\documentclass[aps,prd,a4paper,showpacs,twocolumn]{revtex4-1}
\usepackage{amsmath,amssymb,bm}
\usepackage{gensymb}\usepackage{nicefrac} 
\usepackage{epsfig}
\usepackage{graphicx}
\usepackage{slashed}
\usepackage{epsfig} \usepackage{graphicx} \usepackage{color}
\usepackage{mathrsfs} \usepackage{amssymb} \usepackage{amsmath} \usepackage{url}

\usepackage{xspace} 
\usepackage[bookmarks, breaklinks, colorlinks,urlcolor=black, citecolor=red, 
linkcolor=blue]{hyperref}

\def\jpsi{\ensuremath{{J\mskip -3mu/\mskip -2mu\psi\mskip 2mu}}}
\def\dsp{\displaystyle}
\def\be {\begin{equation}}
\def\ee {\end{equation}}
\def\bea {\begin{eqnarray}}
\def\eea {\end{eqnarray}}
\def\bc {\begin{center}}
\def\ec {\end{center}}
\def\nn {\nonumber}
\def\sss{\scriptscriptstyle}
\def\gev{\ensuremath{\mathrm{Ge\kern -0.1em V}}}

\def \Re{\text{Re}}
\def \Im{\text{Im}}

\def \kstar{{K^{\!*}}}

\def \braket#1#2#3{\langle #1|#2| #3\rangle}
\def\AFB{A_{\text{FB}}}

\def\hel#1{{\sss{#1}}}

\def\AFB{A_{\text{FB}}}

\def\Gf{\Gamma_{\!\! f}}

\def\fb   {\ensuremath{\mbox{\,fb}}\xspace}
\def\invfb   {\ensuremath{\mbox{\,fb}^{-1}}\xspace}
\def\lhcb{LHCb~}

\def\ket|#1>{\left|#1 \right>}

\def\bra<#1|{\left< #1 \right|}

\def\bracket<#1|#2>{\setbox0=\vbox{\hbox{$#1$$#2$}}\left<#1\kern1pt
	\vrule  height\ht0\kern2pt #2\right>} 

\def\dirmat<#1|#2|#3>{\setbox0=\vbox{\hbox{$#1$$#2$$#3$}}\left<#1\kern1pt
	\vrule height\ht0\kern1pt#2\kern1pt \vrule height\ht0\kern1pt
	#3\right>}

\allowdisplaybreaks

\begin{document}

\title{Implications from \texorpdfstring{\boldmath${B\to\kstar\ell^+\ell^-}$}{} 
observables using $3\fb^{-1}$ of \lhcb data.}

\author{Rusa Mandal}\email{rusam@imsc.res.in} 
\affiliation{The Institute of Mathematical
	Sciences, Taramani, Chennai 600113, India \\ and \\ Homi Bhabha National Institute Training School Complex, \\ Anushakti Nagar, Mumbai 400085, India} 
\author{Rahul Sinha}\email{sinha@imsc.res.in}\affiliation{The Institute of 
	Mathematical Sciences, 
	Taramani, Chennai 600113, India \\ and \\ Homi Bhabha National Institute Training School Complex, \\ Anushakti Nagar, Mumbai 400085, India}

\date{\today}

\begin{abstract}
The decay mode $B\to\kstar\ell^+\ell^-$ results in the measurement of a large
number of related observables by studying the angular distribution of the decay
products and is regarded as a sensitive probe of physics beyond the standard
model (SM). Recently, \lhcb has measured several of these observables using
$3\fb^{-1}$ data, as a binned function of $q^2$, the dilepton invariant mass
squared. We show how data can be used without any approximations to extract
theoretical parameters describing the decay and to obtain a relation amongst
observables within the SM. We find three kinds of significant
disagreement between theoretical expectations and values obtained by fits. The
values of the form factors obtained from experimental data show significant
discrepancies when compared with  theoretical expectations in several $q^2$ bins.
We emphasize that this discrepancy cannot arise completely due to resonances and
non-factorizable contributions from charm loops. Further, a relation between
form factors expected to hold at large $q^2$ is very significantly violated.
Finally, the relation between observables also indicates some deviations in the
forward-backward asymmetry in the same $q^2$ regions. 
These discrepancies are possible evidence of physics beyond the SM.
\end{abstract}


\maketitle

\section{Introduction} 
\label{sec:Intro} 

The rare decay $B\to \kstar\ell^+\ell^-$ involves a $b\to s$ flavor changing
loop induced transition at the quark level making it attractive mode to search
for physics beyond the standard model (SM). Indirect searches for new physics 
(NP)
involving loop processes require a comparison of theoretical estimates with
experimental observations. The theoretical estimates thus need to be extremely
reliable in order to make a conclusive claim on the existence or non-existence
of NP. Fortunately, significant progress has been made in understanding the
hadronic effects involved in the decay $B\to \kstar\ell^+\ell^-$. The mode $B\to
\kstar\ell^+\ell^-$ is also of special interest as it allows for the measurement
of several observables using the angular distribution~\cite{Kruger:1999xa}. The 
large number of observables depend on theoretical parameters that describe this 
decay. In this paper we show how some of the parameters can be extracted 
directly from \lhcb measurements allowing us to verify our theoretical 
understanding. Any discrepancy observed must be attributed either to a failure 
of our understanding of hadronic effects or to the existence of NP. We also 
test the relation between observables that provides another clean test for NP.

The differential decay
distribution~\cite{Kruger:1999xa,Das:2012kz,Mandal:2014kma} of $B\to
\kstar\ell^+\ell^-$ results in the measurement of at least nine observables
using the angular distribution, as a function of $q^2$ the dilepton invariant
mass squared. These observables are commonly chosen to be the differential decay
rate with respect to $q^2$, two independent helicity fractions that describe the
decay, the three asymmetries that describe the real part of the interference
between different helicity amplitudes and three asymmetries that describe the
imaginary part of the interference. 

Recently \lhcb~\cite{LHCb:2015dla} has reported measurements of all these
observables that have been averaged in eight $q^2$ bins. A lot of studies on this decay mode are widely discussed in literature \cite{Kruger:2005ep,Descotes-Genon:2015uva}. We use the \lhcb data to obtain estimates of hadronic form factors that describe the decay. Previously
some of the form factors have been determined~\cite{Hambrock:2012dg} using
$1\invfb$ of \lhcb data. We emphasize that our approach does not involve
evaluating the decay amplitude in terms of theoretically estimated parameters.
Instead we start with the most general parametric form of the amplitude without
any hadronic approximations within the SM (see Eq.~\eqref{eq:amp-def2} below).
Experimental data alone is used to fit the theoretical parameters introduced in
the parametric amplitude. These experimentally fitted theoretical parameters are
simply compared to the estimates by other
authors~\cite{Straub:2015ica,Horgan:2013pva} which are widely regarded as the
state of the art.  The values of form factors obtained from experimental
data show significant discrepancy when compared with  theoretical expectations
in several $q^2$ bins.

 In addition to
extracting form factors from data, this mode also allows a relation among
observables that can provide a clean signal~\cite{Mandal:2014kma,Das:2012kz} of
NP. We find that the measurements do not satisfy the expected  relation between
the observables in the same $q^2$ domains where the fitted form factors also
show a large discrepancy with the theoretical estimates. The simultaneous
observation of these discrepancies points to possible evidence of NP.

The paper is organized as follows. In Sec.~\ref{sec:Theory} we describe the theoretical framework developed to write the most general parametric form of the amplitude and cast the observables in a form where hadronic paremeters can be obtainable from data. The relation among observables are also derived here. A numerical analysis is presented in Sec.~\ref{sec:numerics} which contains two subsections. The Sub-sec.~\ref{subsec:ff} gives elaborate description of extraction of form factors using \lhcb measurements, whereas, the validity of the relations derived assuming SM are examined in  Sub-sec.~\ref{subsec:obs} with experimental data. In Sec.~\ref{sec:conclusion} we summarize the important results obtained in this paper. Appendix.~\ref{sec:App1} and \ref{sec:App2} estimate the complex part the amplitude and the systematic uncertainty arising mainly due to bin average effect of the observables, respectively.

\section{Observables and Theoretical Framework} 
\label{sec:Theory} 

In this section we briefly discuss the theoretical framework derived to take into account all possible contributions within SM for the decay $B\to \kstar\ell^+\ell^-$.
We start with the observables as defined in Ref.~\cite{Mandal:2014kma} to be the
$F_L$, $F_\perp$, $A_4$, $A_5$,  $\AFB$, $A_7$, $A_8$, $A_9$ and $d\Gamma/dq^2
\equiv \Gf$. The observables $F_\perp$, $A_4$, $A_5$, $\AFB$, $A_7$, $A_8$ and
$A_9$ are related to the $CP$ averaged observables $S_3$, $S_4$, $S_5$,
$\AFB^{\sss \text{LHC}b }$, $S_7$, $S_8$ and $S_9$ measured by LHCb
\cite{LHCb:2015dla,Aaij:2013iag} as follows:
\begin{align}
\label{eq:S4-S5-S9} F_\perp&=\frac{1-F_L+2 S_3}{2}, ~~A_4=-\dsp\frac{2}{\pi}S_4,
~~A_5=\dsp\frac{3}{4}S_5, \nn \\ 
\AFB\!&=\!-\AFB^{\sss \text{LHCb}}\!,~A_7\!=\!\dsp\frac{3}{4} 
S_7,~A_8\!=-\dsp\frac{2}{\pi}
S_8,~A_9\!=\!\dsp\frac{3}{2\pi}S_9.
\end{align}
It may be remarked that \lhcb collaboration observes a local tension with
some observables based on the hadronic estimates of 
Refs.~\cite{Straub:2015ica,Descotes-Genon:2014uoa}.

We begin by assuming the massless lepton limit but generalize to include the 
lepton mass. The corrections due to the 
mass of the leptons are easily taken into account ~\cite{Mandal:2014kma}.
 In the massless lepton limit the decay is described in terms of 
six transversity amplitudes which can be written in the most general form as,
\begin{equation}
\mathcal{A}_\lambda^{L,R}=C_{L,R}^{\sss\lambda}\,\mathcal{F}_\lambda 
-\widetilde{\mathcal{G}}_\lambda
\label{eq:amp-def1}
=\big(\widetilde{C}_9^{\sss\lambda}\mp 
C_{10})\mathcal{F}_\lambda -\widetilde{\mathcal{G}}_\lambda.
\end{equation}
This form of
the amplitude~\cite{Mandal:2014kma} is the most
general parametric form of SM amplitude for $B\to
\kstar\ell^+\ell^-$ decay that  comprehensively takes into account all 
contributions up to $\mathcal{O}(G_{\!F})$
within it. The form includes all short-distance and long-distance
effects, factorizable and nonfactorizable contributions and resonance contributions. 
In Eq.~\eqref{eq:amp-def1} $C_9$ and $C_{10}$ are Wilson 
coefficients with 
$\widetilde{C}_9^{\sss\lambda}$ being the redefined ``effective'' Wilson 
coefficient defined~\cite{Mandal:2014kma,Beneke:2001at} such that 
\begin{equation}
\widetilde{C}_9^{\sss\lambda}=C_9+\Delta 
C_9^{\text{(fac)}}(q^2)+\Delta C_9^{{\sss\lambda}{\text{,(non-fac)}}}(q^2)
\end{equation}
where $\Delta C_9^{\text{(fac)}}(q^2)$, $\Delta 
C_9^{{\sss\lambda}{\text{,(non-fac)}}}(q^2)$
correspond to factorizable and soft gluon non-factorizable contributions.
Strong interaction effects coming from electromagnetic
corrections to hadronic operators do not affect $C_{10}$.

The form factors $\mathcal{F}_\lambda$ and $\widetilde{\mathcal{G}}_\lambda$
introduced in Eq.~\eqref{eq:amp-def1} 
can be related to the conventional
form factors describing the decay as shown in the appendix of
Ref.~\cite{Mandal:2014kma}. The form-factors $\mathcal{F}_\lambda$ are of
particular interest here as we show that they can be extracted directly from
data. The  $\mathcal{F}_\lambda$ can be related to the well known form-factors
$V$, $A_1$ and $A_{12}$ by comparing with \cite{Straub:2015ica}:
\begin{subequations}
\begin{align}
\label{eq:V}
\mathcal{F}_\perp=&  N\sqrt{2}\sqrt{\lambda(m_B^2,m_\kstar^2,q^2)}
\frac{V(q^2)}{m_B + m_{K^*}}, \\
\label{eq:A1}
\mathcal{F}_\|= &-N \sqrt{2}(m_B + m_{K^*}) A_1(q^2), \\
\label{eq:A12}
\mathcal{F}_0=& \frac{-N}{\sqrt{q^2}}\, 8 m_B m_\kstar A_{12}(q^2).
\end{align}
\end{subequations}

 It should be noted that
$\mathcal{F}_\lambda$'s and $C_{10}$ are completely real in the SM, with all
imaginary contributions to the amplitude arising only from the imaginary part of complex
$\widetilde{C}_9^{\sss\lambda}$ and $\widetilde{\mathcal{G}}_\lambda$ terms.
Thus with the introduction of two variables $r_\lambda$ and $\varepsilon_\lambda$
the amplitude ${\cal A}_\lambda^{L,R}$ in 
Eq.~\eqref{eq:amp-def1} can be rewritten as,
\begin{equation}
\label{eq:amp-def2}
\mathcal{A}_\lambda^{L,R}=(\mp 
C_{10}-r_\lambda)\mathcal{F}_\lambda+i\varepsilon_\lambda,
\end{equation}
where, \vspace*{-0.5cm}
\begin{eqnarray}
\label{eq:rlambda}
r_\lambda&=&\frac{\Re(\widetilde{\mathcal{G}}_\lambda)}{\mathcal{F}_\lambda}
-\Re(\widetilde{C}_9^\hel{\lambda}),\\
\label{eq:epsilon}
\varepsilon_\lambda &=&
\Im(\widetilde{C}_9^\hel{\lambda})\mathcal{F}_\lambda
-\Im(\widetilde{\mathcal{G}}_\lambda).
\end{eqnarray}
The imaginary contributions arise mostly from resonant long-distance contributions,
which can be removed by studying only those $q^2$ regions where no resonances can contribute.
In practice this means the removal of charmonium resonance regions from the whole $q^2$ range. 
\lhcb $3\fb^{-1}$ measurements~\cite{LHCb:2015dla} conservatively
exclude the resonance region.
Moreover, the contributions from imaginary parts are bounded directly from
the \lhcb measurements and the bin average values of the $\varepsilon_\lambda$'s 
are found to be very small as shown in Appendix.~\ref{sec:App1}. Hence for now we are neglecting the $\varepsilon_\lambda$'s and will address it's contribution in the numerical analysis.
\begin{center}
\begin{figure*}[htbp]
 \begin{center}
	\includegraphics*[width=1.6in]{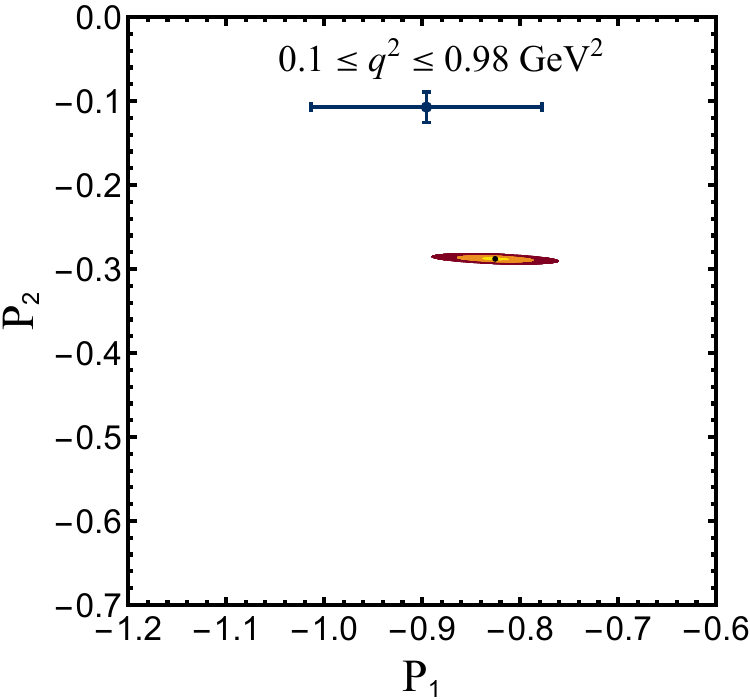}%
	\includegraphics*[width=1.6in]{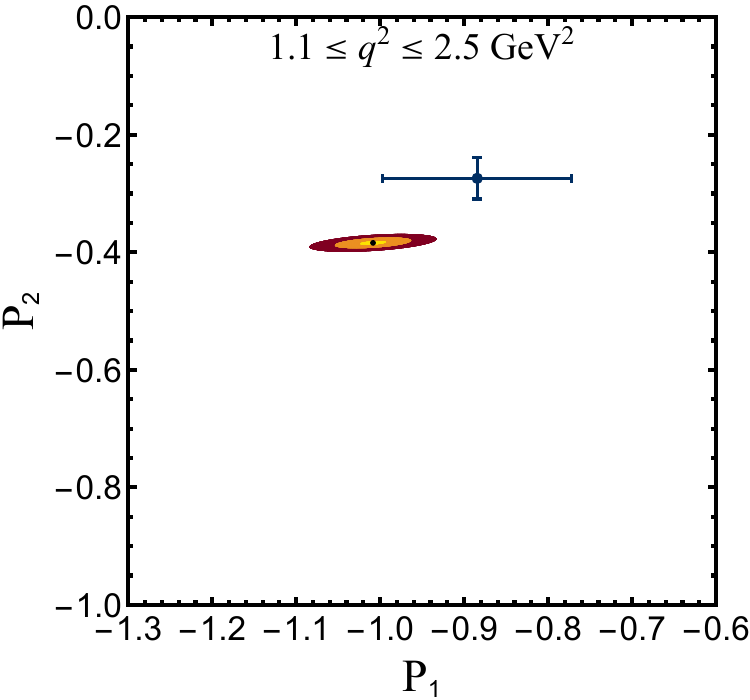}
	\includegraphics*[width=1.6in]{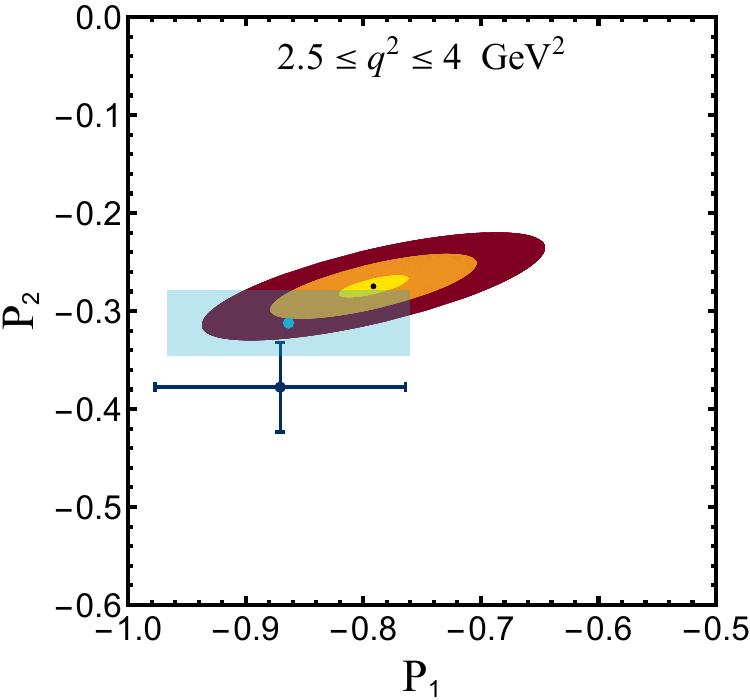}
	\includegraphics*[width=1.6in]{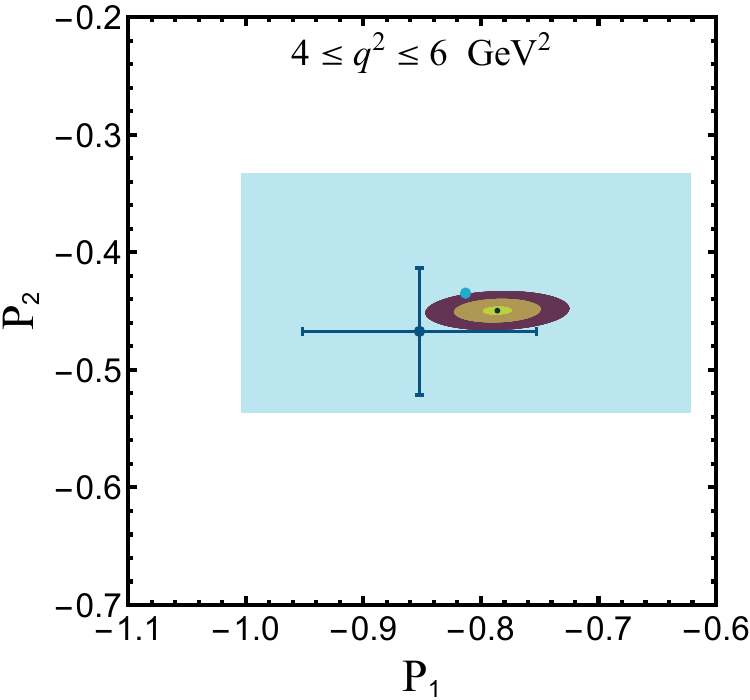}
	\includegraphics*[width=1.6in]{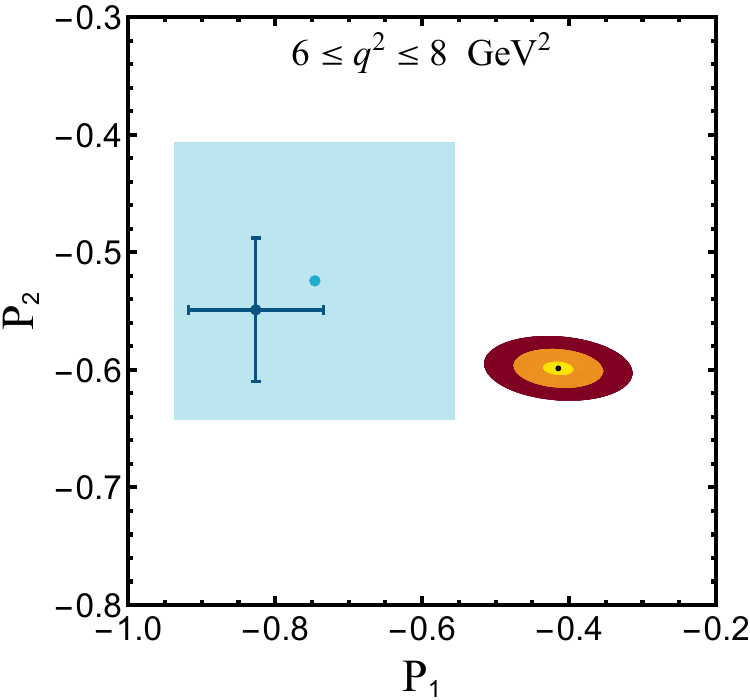}%
	\includegraphics*[width=1.6in]{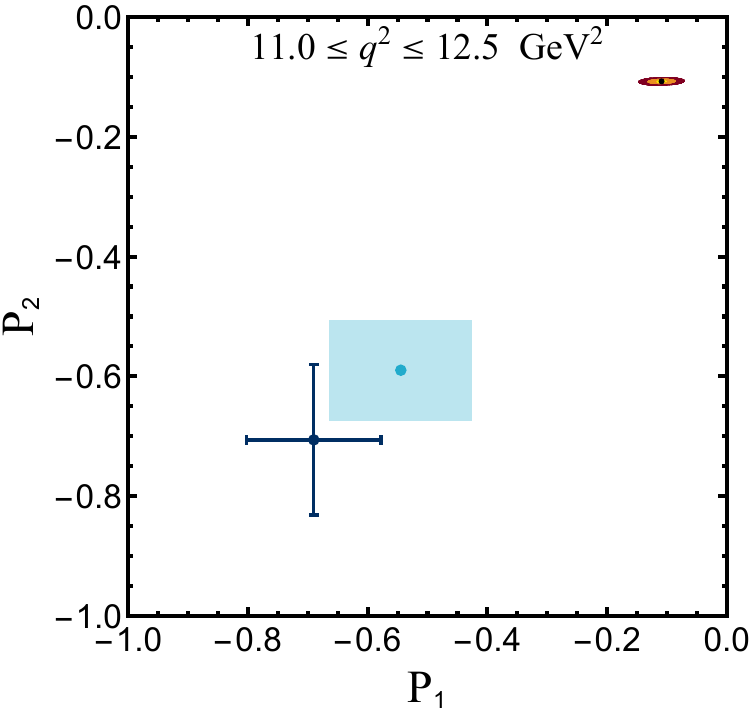}
	\includegraphics*[width=1.6in]{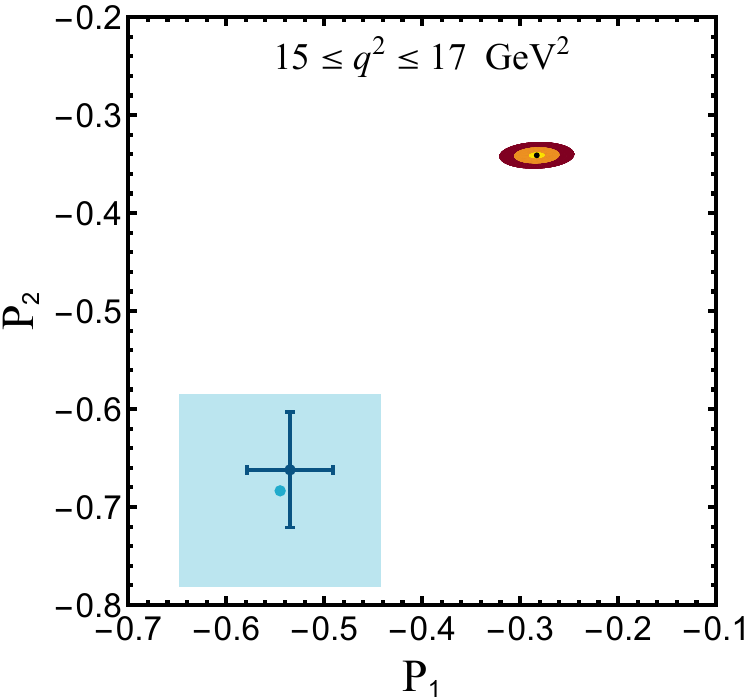}
	\includegraphics*[width=1.6in]{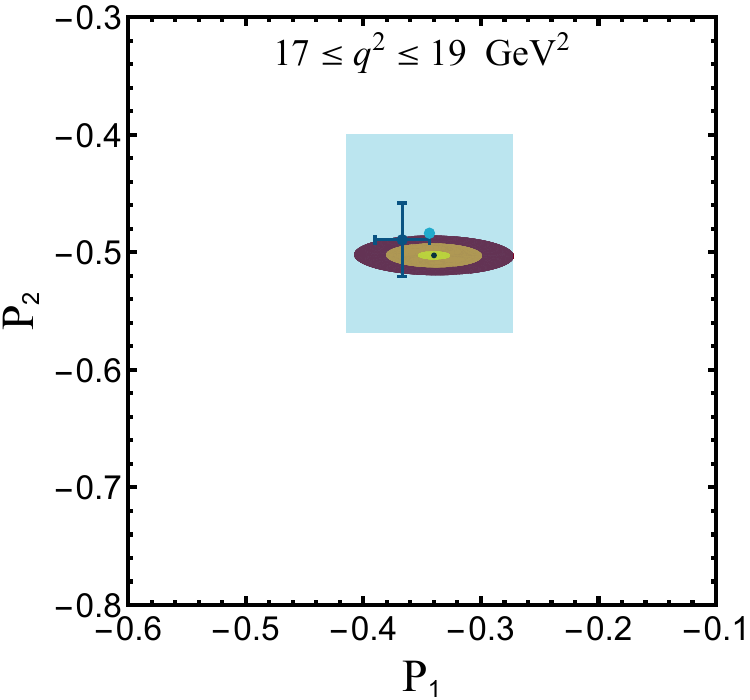} \caption{(color online). 
	The allowed region for $\mathsf{P}_1$ versus $\mathsf{P}_2$ plane. The 
	innermost yellow (lightest), the middle one orange (light) and outer most 
	red (dark) contours represent $1\sigma$, $3\sigma$ and $5\sigma$ regions, 
	respectively. The theoretically estimated values using 
	Ref.~\cite{Straub:2015ica} for 
	$q^2\le 8~\gev^2$ and Ref.~\cite{Horgan:2013pva} for $q^2\ge 11~\gev^2$ 
	are shown as points with error bars. The light blue bands denote exact solutions for the SM observables including charmonium resonances from Ref.~\cite{Kruger:1996cv} parametrization and are shown only for the relevant $q^2$ bins. In most cases, there is 
	reasonable agreement between the theoretical values and those obtained 
	from data. However, for the ranges $0.1\le q^2 \le 0.98 \gev^2$, $6\le q^2 \le 	    
	8\gev^2$, $11.0\le q^2\le 12.5 ~\gev^2$ and $15\le q^2\le 17~\gev^2$ there are 
	significant disagreements.} \label{fig:P1P2}
 \end{center}
\end{figure*}
\end{center}

\vspace*{-0.5cm}
It is convenient to define
$\mathsf{P}_1$ and $ \mathsf{P}_2$ as, 
\begin{equation}
  \label{eq:P_1-2}
  \mathsf{P}_1=\frac{\mathcal{F}_\perp}{\mathcal{F}_\|},\qquad  
  \mathsf{P}_2=\dsp \frac{\mathcal{F}_\perp}{\mathcal{F}_0}.
\end{equation}
The observables $F_\perp$, $F_L$, $\AFB$, $A_5$ and $A_4$ can be
written~\cite{Mandal:2014kma} as
\begin{align}
\label{eq:Fperp} 
F_{\!\perp} &=u_{\!\perp}^2 +2 \zeta\\
\label{eq:Fparallel} 
F_L \mathsf{P_2^2} &= u_0^2+ 2\zeta\\ 
\label{eq:AFB} 
\AFB^2 &=\frac{9\zeta}{2\mathsf{P}_1^2}
\big(u_\|\pm u_\perp \big)^2\\
\label{eq:A5} 
A_5^2 &=\frac{9\zeta}{4\mathsf{P}_2^2}
\big(u_0 \pm u_\perp \big)^2\\
\label{eq:A4} 
A_4 &=\frac{\sqrt{2}}{\pi \mathsf{P}_1\mathsf{P}_2}
\big(2\zeta \pm u_0 u_\| \big)
\end{align}
where, 
\vspace*{-0.2cm}
\begin{align}
\label{eq:zeta} & 
\zeta = \frac{\mathcal{F}_\perp^2 C_{10}^2}{\Gf}, \\
& 
u_\lambda^2\!=\!\frac{2\mathcal{F}_\perp^2 r_\lambda^2}{\Gf}\!=\!
\frac{2}{\Gf}\frac{\mathcal{F}_\perp^2}{\mathcal{F}_\lambda^2}
\Big(\Re(\widetilde{\mathcal{G}}_\lambda)-\Re(\widetilde{C}_9^\hel{\lambda})
\mathcal{F}_\lambda \Big)^2.
\end{align}
$u_\lambda$ is always taken to be positive and the sign ambiguities introduced
in Eqs.~\eqref{eq:AFB}-\eqref{eq:A4} ensure that we can make this assumption.
The five observables $F_\perp$, $F_L$, $\AFB$, $A_5$ and $A_4$  have been
expressed above in terms of five parameters $\mathsf{P}_1$, $\mathsf{P}_2$,
$\zeta$, $u_0$ and $u_\perp$. The other three observables $A_7$, $A_8$ and $A_9$
have already been used to solve for the three $\varepsilon_\lambda$ values
which are presented in Ref.~\cite{Mandal:2014kma}. It may be noted that since
$F_\|=1-F_L-F_\perp$, $u_\|$ is not independent and is related to the other
parameters by, $u_\|^2= \mathsf{P}_1^2 \big( 
1-\mathsf{P}_2^{-2}(u_0^2+2\zeta)-(u_\perp^2+2\zeta)\big)-2\zeta$.

In Refs.~\cite{Mandal:2014kma,Das:2012kz} a relation depending on observables including all possible effects within SM which was derived as,
\begin{widetext}
\begin{equation}
\sqrt{4\Big( F_L \!+\! F_\| \!+ \!\sqrt{2}\pi A_4 \Big) F_\perp \!-\! 
\frac{16}{9} \Big(\AFB \!+\! \sqrt{2}A_5\Big)^2}=\sqrt{4 F_\|  
F_\perp\!-\!\frac{16}{9}\AFB^2}+\sqrt{4 F_L F_\perp\!-\!\frac{32}{9}A_5^2}.
\end{equation}
\end{widetext}
This equation can be used to express any of the observables in terms of the 
others. A solution for $A_4$ ~\cite{Das:2012kz} is
\begin{equation}
  \label{eq:Obs-relationA4} \!\!A_4\!=\! \frac{8 A_5 \AFB}{9 \pi F_\perp}
  \!+\!\frac{\sqrt{4 F_\|  F_\perp\!-\!\frac{16}{9}\AFB^2} \sqrt{4 F_L
  F_\perp\!-\!\frac{32}{9}A_5^2}}{2\sqrt{2}\pi F_\perp}.
\end{equation}
Whereas, the solution for $A_5$ and $\AFB$ are given by,
\begin{align}
\label{eq:Obs-relationA5}
\!\!& A_5 \!=\!\! \frac{\pi A_4 \AFB}{2F_\|}\! \pm\! \frac{3\sqrt{\! 4 F_\|  
F_\perp\!\!-\!\frac{16}{9}\AFB^2}\!\sqrt{\!\,2 F_\| F_L\!\!-\!\pi^2\! A_4^2}}{8 F_\|}, \\
\label{eq:Obs-relationAFB}
\!\!& \AFB \!=\!\! \frac{\pi A_4 A_5}{F_L}\! \pm\! \frac{3\sqrt{\!4 F_L 
F_\perp\!\!-\!\frac{32}{9} A_5^2}\sqrt{ 2 F_\| F_L\!-\!\pi ^2 A_4^2}}{4\sqrt{2} F_L}.
\end{align}
{\em It may noted that Eqs.~\eqref{eq:Obs-relationA4}, \eqref{eq:Obs-relationA5} and 
\eqref{eq:Obs-relationAFB} depend only on observables and not on any theoretical 
parameters and thus provides an exact test of the gauge structure of SM with experimental measurements.}

\begin{center}
\begin{figure*}[hbtp]
 \begin{center}
	\includegraphics*[width=1.6in]{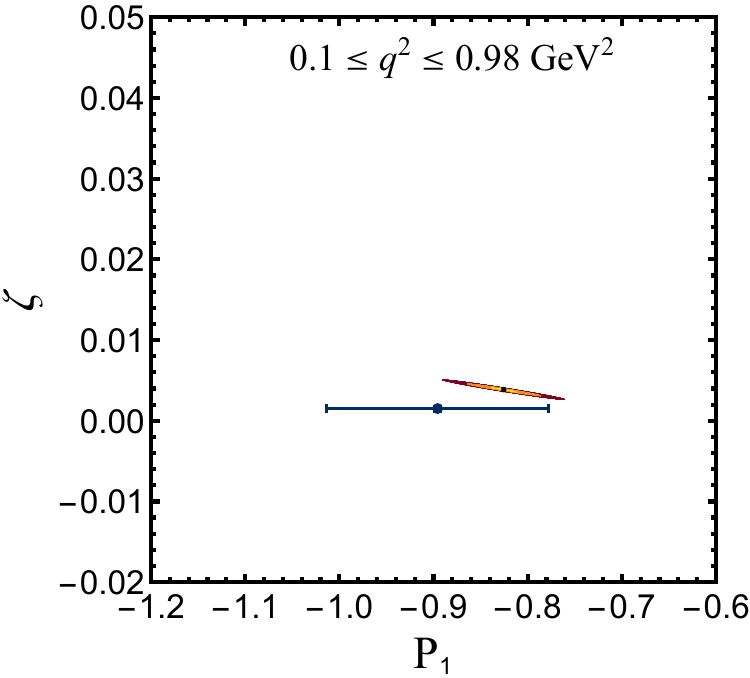}%
	\includegraphics*[width=1.6in]{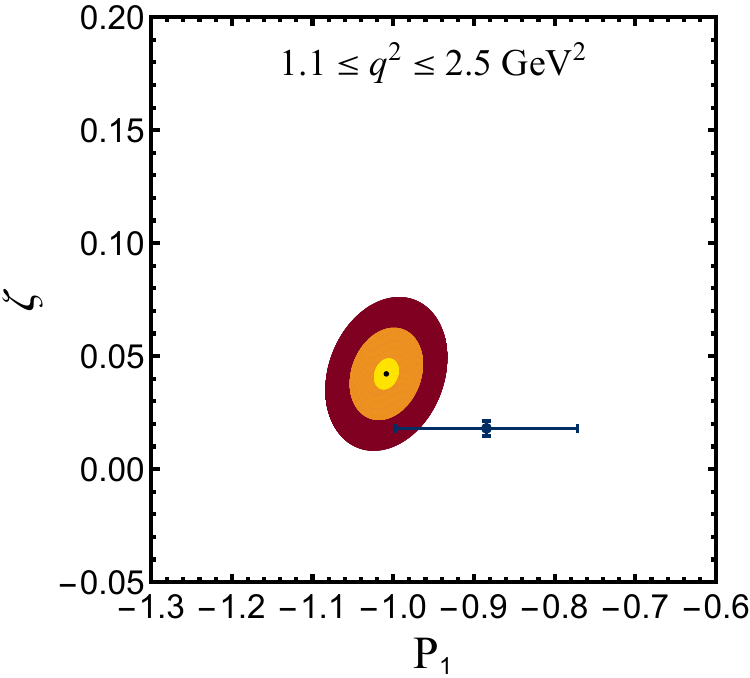}
	\includegraphics*[width=1.6in]{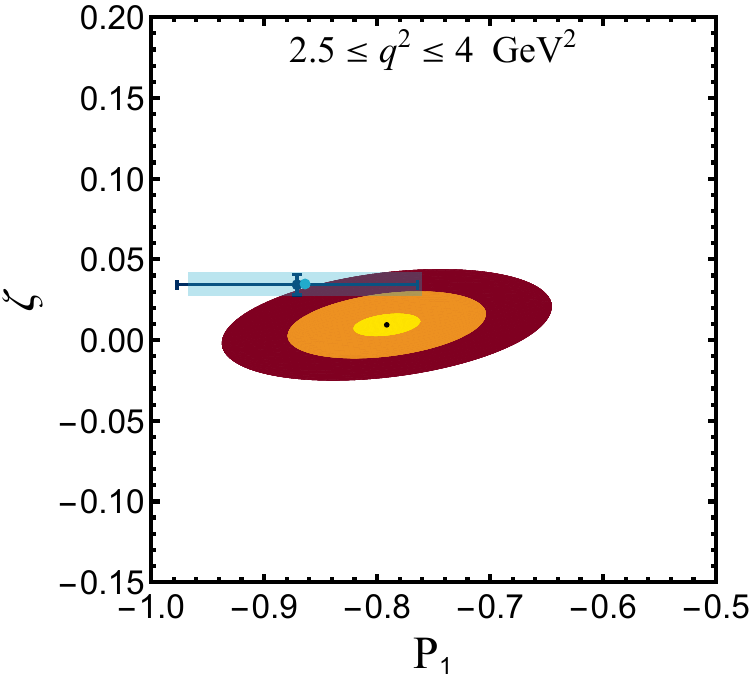}
	\includegraphics*[width=1.6in]{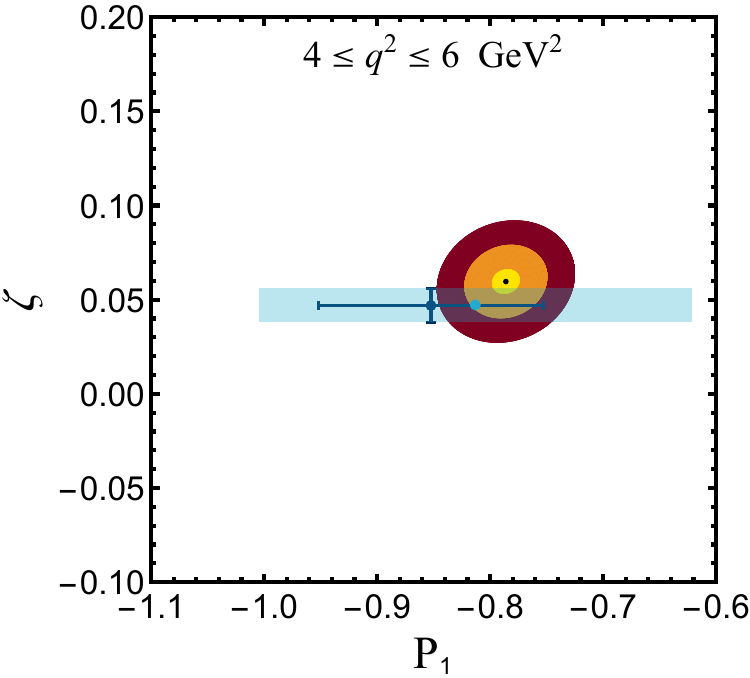}
	\includegraphics*[width=1.6in]{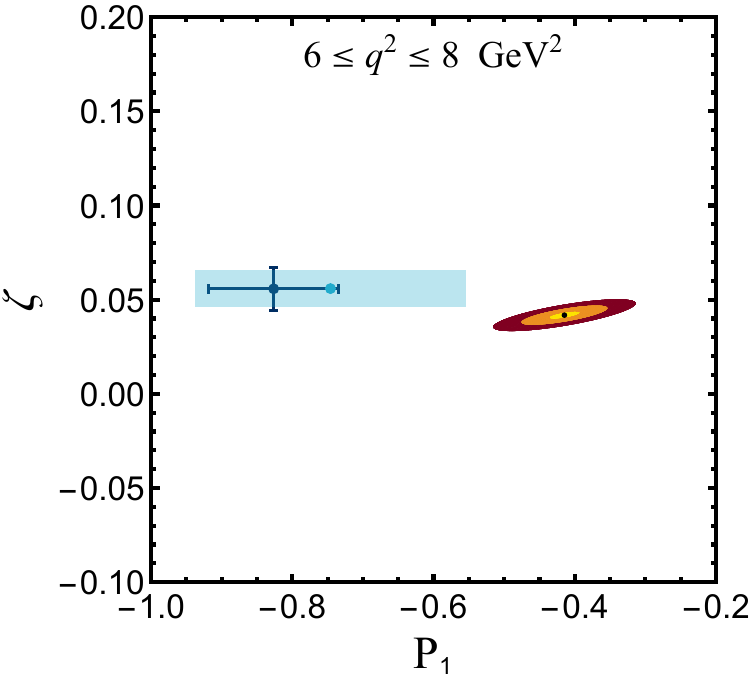}%
	\includegraphics*[width=1.6in]{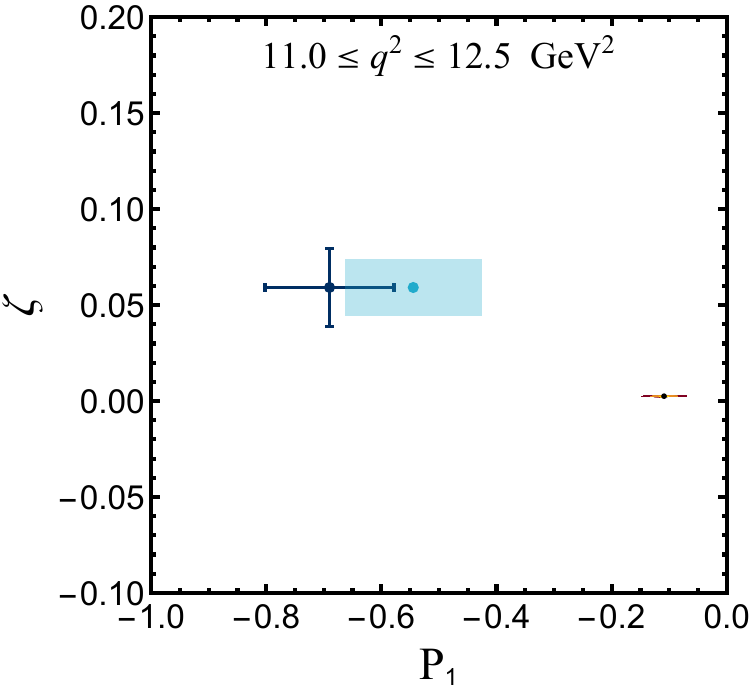}
	\includegraphics*[width=1.6in]{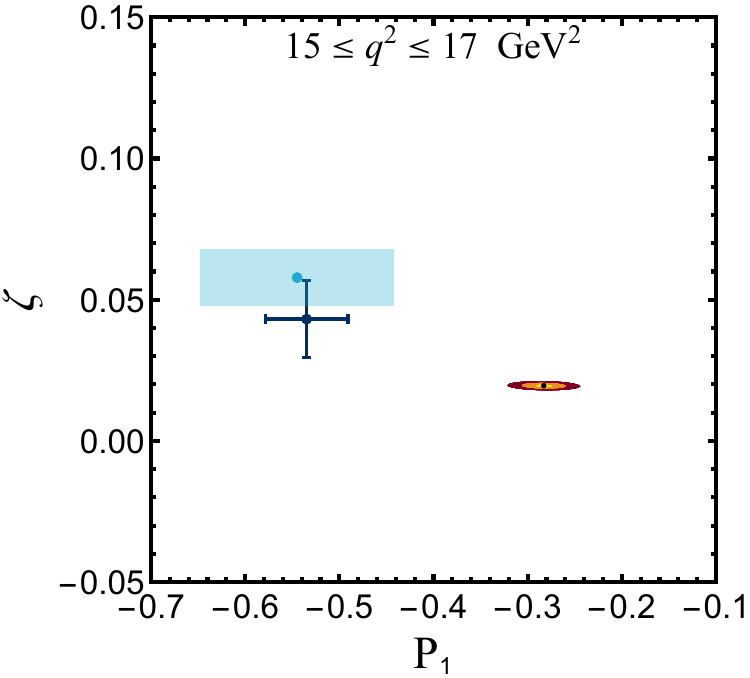}
	\includegraphics*[width=1.6in]{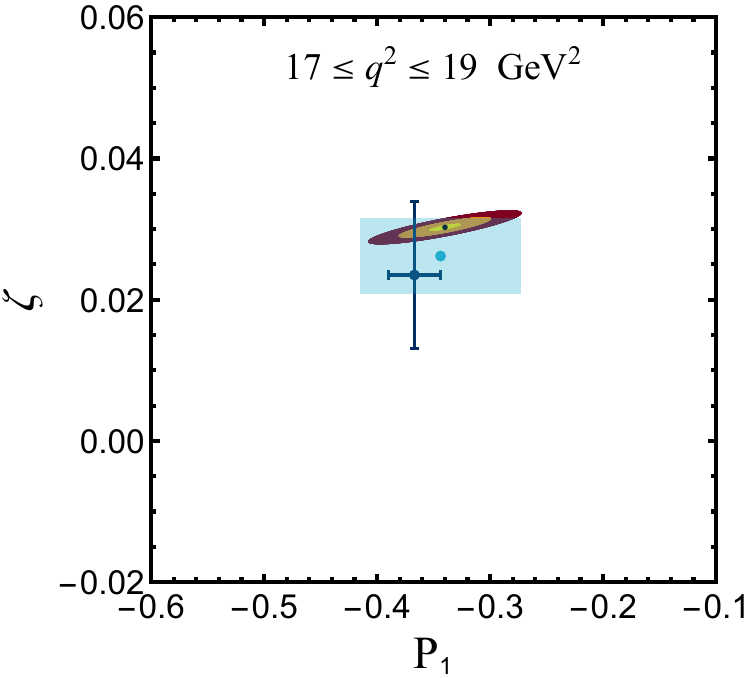} \caption{(color online). The
	allowed region for $\mathsf{P}_1$ versus $\mathsf{\zeta}$ plane. The 
	color code is same as Fig.~\ref{fig:P1P2}. The theoretically estimated 
	values from Ref.~\cite{Straub:2015ica,Horgan:2013pva}
	are shown as points with error bars. The $\mathsf{P}_1$ and $\mathsf{\zeta}$ values 
	significantly disagree for ranges $6\le q^2 \le 8\gev^2$,  $11.0\le q^2\le 12.5 
	~\gev^2$ and $15\le q^2\le 17 ~\gev^2$, similar to the values of $\mathsf{P}_1$ and 
	$\mathsf{P}_2$ shown in Fig.~\ref{fig:P1P2}.} 
	\label{fig:P1zeta}
\end{center}
\end{figure*}
\end{center}

\section{New Physics Analysis} 
\label{sec:numerics} 
In this section we discuss the detailed numerical analysis using $3\invfb$ of \lhcb data \cite{LHCb:2015dla}. It contains two different parts, at first we show how the experimental data can be used to extract out the form factors which are involved in this decay mode. Secondly we present the consistency of data to test the relation among observables derived relying only on the gauge structure of SM. 

\subsection{Form Factor Extraction}
\label{subsec:ff} 

We demonstrate the technique to extract out the hadronic parameters by including complex contributions of the amplitudes and considering systematic uncertainty arising mainly due to bin average effect.

It is shown in Ref.~\cite{Mandal:2014kma} that $\varepsilon_\lambda$'s contribute to the helicity fractions $F_\lambda$ and asymmetry $A_4$. We refer to Appendix.~\ref{sec:App1} for thorough details of evaluation of the complex part of the amplitudes. Using the bin average central values of $\varepsilon_\lambda/\sqrt{\Gf}$, with $\pm 1 \sigma$ errors from Table.~\ref{eps-Table} we can numerically separate out the complex contributions from experimental measured values of the observables. We calculate the central value with $\pm 1 \sigma$ error of the modified observables ${F_\lambda^\text{ex}}^\prime$ and ${A_4^\text{ex}}^\prime$ given by,
\begin{align}
\label{eq:FLprime}
{F_\lambda^\text{ex}}^\prime&=F_\lambda^\text{ex}-\frac{2 \varepsilon_\lambda^2}{\Gf}, \\
\label{eq:A4prime}
{A_4^\text{ex}}^\prime&=A_4^\text{ex}-\frac{2\sqrt{2}\varepsilon_0\varepsilon_\|}{\pi\Gf},
\end{align}
which enter in the $\chi^2$ definition Eq.~\eqref{eq:chisq} below. It enables us to take into account the complex corrections in our analysis and extract out the variables $\mathsf{P}_1$,
$\mathsf{P}_2$, $\zeta$, $u_0$ and $u_\perp$ (which only deal with the real part of amplitude) from experimental measurements of the observables accurately.

It should be noted that Eqs.~\eqref{eq:Fperp} -- \eqref{eq:A4} are valid for each $q^2$ point. However, experiments can provide bin integrated values of observables over a certain $q^2$ intervals. Thus a $\chi^2$ fit with bin average values of the observables may lead to a biased conclusion. To avoid this issue we have added systematic uncertainties for each observables due to bin average effect with the introduction of new parameter $\beta$, where the change in each observable $\mathcal{O}$ is given by,
\begin{align}
\label{eq:shift}
\mathcal{O} \to \mathcal{O} + \beta\, \mathcal{O}^{s}. 
\end{align}
$\mathcal{O}^{s}$ is the maximum shift for each observables with a best fitted $q^2$ function to 14 bin \lhcb data \cite{LHCb:2015dla} within the concerned bin interval. The precise determination of $\mathcal{O}^{s}$ is described in Appendix.~\ref{sec:App2}.
%
%
%
Therefore the $\chi^2$ definition is
\begin{align}
\label{eq:chisq}
\chi^2&=\Bigg[\Bigg(\frac{{F_\perp^\text{ex}}^\prime -F_\perp - \beta F_\perp^s}{\Delta {F_\perp^\text{ex}}^\prime} \Bigg)^{\!2}
+\Bigg(\frac{{F_L^\text{ex}}^\prime -F_L- \beta F_L^s}{\Delta {F_L^\text{ex}}^\prime}\Bigg)^{\!2} \nn\\&+
\Bigg(\frac{{A_4^\text{ex}}^\prime -A_4- \beta A_4^s}{\Delta {A_4^\text{ex}}^\prime }\Bigg)^{\!2} +
\Bigg(\frac{{\AFB^2}^\text{\!\!\!\!ex}-\AFB^2- \beta {\AFB^2}^{\!\!\!\!\!s}}{2\AFB^\text{ex}\Delta \AFB^\text{ex}}\Bigg)^{\!2} \nn\\&+
\Bigg(\frac{{A_5^2}^\text{ex}-A_5^2- \beta {A_5^2}^s}{2A_5^\text{ex}\Delta A_5^\text{ex}}\Bigg)^{\!2\,} + \beta^2 \Bigg],
\end{align}

where $\,\AFB^\text{ex}$ and $\,A_5^\text{ex}$ indicate experimental central values of
the observables $\AFB$ and $A_5$ with $\pm 1 \sigma$ errors as $\Delta\AFB^\text{ex}$ and $ \Delta A_5^\text{ex}$, respectively. Similarly $\,{F_\perp^\text{ex}}^\prime$, $\,{F_L^\text{ex}}^\prime$ and $\,{A_4^\text{ex}}^\prime$ are the central values of the modified observables defined in Eqs.~\eqref{eq:FLprime} and \eqref{eq:A4prime} and $ \Delta {F_\perp^\text{ex}}^\prime, \Delta {F_L^\text{ex}}^\prime, \Delta {A_4^\text{ex}}^\prime$ are $\pm 1 \sigma$ uncertainties in it.
The systematic uncertainties added for each observables are denoted as $F_\perp^s$, $F_L^s$, $A_4^s$, ${\AFB^2}^{\!\!\!\!\!s}$ , ${A_5^2}^{\!\,s}$ and these values are quoted in Table.~\ref{sys-Table} of Appendix.~\ref{sec:App2}.
The observables $\,F_\perp$, $\,F_L$, $\,A_4$, $\,\AFB^2$ and
$A_5^2$ are evaluated in terms of the five parameters $\mathsf{P}_1$,
$\mathsf{P}_2$, $\zeta$, $u_0$ and $u_\perp$ using
Eqs.~\eqref{eq:Fperp} -- \eqref{eq:A4}. Considering the inverse of the covariance
matrix the error ellipsoids are constructed for all the eight bins corresponding
to the $q^2$ values in the range $(0.1-0.98)~\text{GeV}^2$, $(1.1-2.5)~\gev^2$,
$(2.5-4.0)~ \gev^2$, $(4-6)~\gev^2$, $(6-8)~\gev^2$, $(11.0-12.5)~\gev^2$,
$(15-17)~\gev^2$ and $(17-19)~\gev^2$.
It can be seen that $\beta$ is treated as a nuisance parameter with values $0\pm1$. The $\chi^2$ function is minimized w.r.t six parameters $\mathsf{P}_1$,
$\mathsf{P}_2$, $\zeta$, $u_0$, $u_\perp$ and $\beta$ and the contours shown in 
Fig.~\ref{fig:P1P2} and Fig.~\ref{fig:P1zeta} are the allowed regions in the 
corresponding planes. The minimum values of the $\chi^2$ function for first to 
eighth bins are $6.9\times 10^{-9}$, $3.4\times 10^{-10}$, $0.055$, $8.6\times 
10^{-30}$, $1.094$, $0.538$, $0.218$ and $0.044$, respectively.
The best fitted values with $\pm 1 \sigma$ errors of the parameter $\beta$ for all eight bins are $ 7.4 \times 10^{-5}\pm 0.015,~ 1.6 \times 10^{-5} \pm 0.020,~ 0.153\pm 0.011,~ 1.0\times 10^{-17} \pm 0.005,~ 0.736 \pm 0.020,~ 0.251 \pm 0.003,~ 0.261 \pm 0.001 \text{~and~} 0.161 \pm 0.012$, respectively.

The contours corresponding to $1\sigma$, $3\sigma$ and $5\sigma$ permitted
regions for $\mathsf{P}_1$ versus $\mathsf{P}_2$ plane are presented in
Fig.~\ref{fig:P1P2}. These contours are compared with the estimated values of
$\mathsf{P}_1$ and $\mathsf{P}_2$ using Ref.~~\cite{Straub:2015ica} for $q^2\le
8~\gev^2$ and Ref.~\cite{Horgan:2013pva} for $q^2\ge 11~\gev^2$. The center
black point denotes the best fit point by minimizing the chi-square function
defined in Eq.~\eqref{eq:chisq}. In most cases reasonable agreement is found
between theoretical values of $\mathsf{P}_1$ and $\mathsf{P}_2$ and their values
obtained from data. However, there are some significant disagreements. The
values of form factor ratio $\mathsf{P}_2$ differ by $9\sigma$ in the
$0.1\!\le\! q^2\!\le\! 0.98~\gev^2$ bin. It may be noted that this region
in $q^2$ is highly affected by finite lepton mass and hence the large
discrepancy may not accurately reflect the significance due to the unaccounted
lepton mass correction systematics. 
Significant deviations are also found for the three bins $6\le q^2\le 8
~\gev^2$, $11.0\le q^2\le 12.5 ~\gev^2$ and $15\le q^2\le 17 ~\gev^2$ where
$\mathsf{P}_1$ ($\mathsf{P}_2$) differ by $4.2\sigma$ ($0.8\sigma$), $5.2\sigma$
($4.8\sigma$) and $5.5\sigma$ ($5.3\sigma$), respectively. The light blue bands denote exact solutions for the SM observables including charmonium resonances from Ref.~\cite{Kruger:1996cv} parametrization and are shown only for the relevant $q^2$ bins. The detailed analysis of resonance effect will be discussed later in this section.

In Fig.~\ref{fig:P1zeta} contours similar to Fig.~\ref{fig:P1P2}, but
corresponding to $\mathsf{P}_1$ versus $\zeta$ permitted regions are presented
for $1\sigma$, $3\sigma$ and $5\sigma$ confidence level regions. These contours
are similarly compared with the estimated values of $\mathsf{P}_1$ and $\zeta$
using Refs.~\cite{Straub:2015ica,Horgan:2013pva} and 
assuming the theoretical estimate of
$C_{10}$~\cite{Altmannshofer:2008dz}. Data shows consistency with 
theoretical values of $\mathsf{P}_1$ and $\zeta$ in most cases
except for the two bins $11.0\le q^2\le 12.5 ~\gev^2$
and $15\le q^2\le 17 ~\gev^2$ where $\zeta$ disagrees by $2.8\sigma$ and 
$1.7\sigma$ respectively. The best fit value of $\zeta$ with $\pm 1\sigma$
error obtained from the fit can be used to calculate the form factor
$\mathcal{F}_\perp$ using Eq.~\eqref{eq:zeta}. 

\begin{table}[h]
\centering
  \begin{tabular}{ |c|  c |c |c| }
    \hline \hline
    $q^2$ range in $\gev^2$ & $V(q^2)$ & $A_1(q^2)$ & $A_{12}(q^2)$  \\ \hline
    $0.1\le q^2 \le 0.98$ & $0.677 \pm 0.092$ & $0.570 \pm 0.077$ & $0.246 \pm 0.034$
     \vspace*{-0.15cm}\\
    & (3.05$\sigma$) & (3.40$\sigma$)  & (0.88$\sigma$)
    \\    \hline
    $1.1\le q^2 \le 2.5$  & $0.625 \pm 0.071$ & $0.409 \pm 0.046$ & $0.326 \pm 0.047$
    \vspace*{-0.15cm}\\ 
    & (2.78$\sigma$) & (2.00$\sigma$)  & (0.69$\sigma$)
    \\    \hline
    $2.5\le q^2 \le 4.0 $ & $0.230 \pm 0.150$ & $0.180 \pm 0.118$ & $0.214 \pm 0.149$ 
    \vspace*{-0.15cm}\\ 
    & (1.36$\sigma$) & (1.09$\sigma$)  & (0.81$\sigma$)
    \\    \hline
    $4.0\le q^2 \le 6.0 $ & $0.552 \pm 0.043$ & $0.400 \pm 0.032$ & $0.359 \pm 0.041$
    \vspace*{-0.15cm}\\ 
    & (1.07$\sigma$) & (1.69$\sigma$)  & (1.09$\sigma$)
    \\    \hline
    $6.0\le q^2 \le 8.0 $ &$0.485 \pm 0.045$ & $0.598 \pm 0.073$ & $0.252 \pm 0.025$
    \vspace*{-0.15cm}\\ 
    & (1.27$\sigma$) & (3.18$\sigma$)  & (1.78$\sigma$)
    \\    \hline
    $11.0\le q^2 \le 12.5$& $0.166 \pm 0.018$ & $0.560 \pm 0.065$ & $0.450 \pm 0.054$
    \vspace*{-0.15cm}\\ 
    & (5.64$\sigma$) & (1.76$\sigma$)  & (1.81$\sigma$)
    \\    \hline
    $15.0\le q^2 \le 17.0 $ & $0.828 \pm 0.120$ & $0.649 \pm 0.098$ & $0.496 \pm 0.074$
    \vspace*{-0.15cm} \\ 
    & (2.79$\sigma$) & (1.38$\sigma$)  & (1.51$\sigma$)
    \\    \hline
    $17.0\le q^2 \le 19.0 $& $1.813 \pm 0.436$ & $0.698 \pm 0.171$ & $0.461 \pm 0.112$
    \vspace*{-0.15cm}\\
    & (0.78$\sigma$) & (0.80$\sigma$)  & (0.91$\sigma$)
    \\  
    \hline
    \hline 
  \end{tabular}
  \caption{The form factor values obtained from
  fit to $3\fb^{-1}$ of \lhcb data \cite{LHCb:2015dla}. Round brackets indicate 
  the standard deviation
   between fitted values and theoretical
  estimates~\cite{Straub:2015ica, Horgan:2013pva}. Significant
  discrepancies are found for $V$ and $A_1$ in several $q^2$ region.
}  \label{Table}
\end{table}

\begin{center}
	\begin{figure*}[t]
		\begin{center}
			\includegraphics*[width=3in]{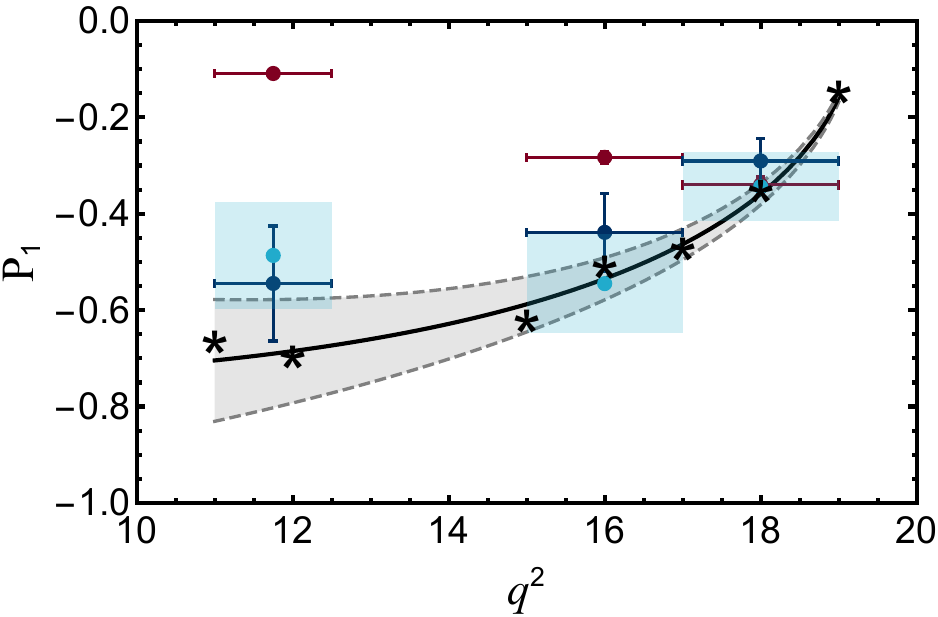}%
			\includegraphics*[width=3in]{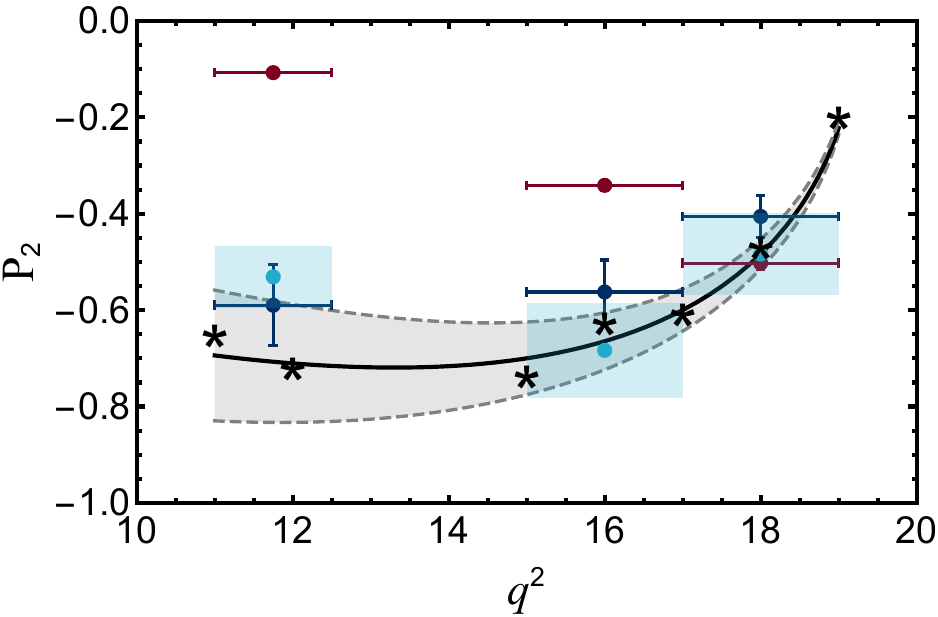}%
			\caption{(color online) Illustrative plots for bin average and resonance effects in the solutions for $\mathsf{P_1}$ (left panel) and $\mathsf{P_2}$ (right panel). The SM observables are assumed from lattice form factors \cite{Horgan:2013pva}. The black `stars' denote the solutions obtained at seven different points in $q^2$ for the corresponding parameters in each plot. The black central curve with gray band is the form factor estimate (mean with $\pm 1\sigma$ error) of $\mathsf{P_1}$ and $\mathsf{P_2}$. The blue error bars are the solutions for $\mathsf{P_1}$ and $\mathsf{P_2}$ using the bin average values of SM observables whereas the light blue bands denote the solutions considering resonances in observables from Ref.~\cite{Kruger:1996cv} parametrization. The red error bars denote the solutions obtained using data (as highlighted in contours is Fig.~\ref{fig:P1P2}). Including the resonances with the parametrization used in Ref.~\cite{Kruger:1996cv}, the solutions for $\mathsf{P_1}$ and $\mathsf{P_2}$ are unaltered and superimpose with the `stars' completely. (see text for details)} 
			\label{fig:P1-P2-Check}
		\end{center}
	\end{figure*}
\end{center}

Finally the form factor $V$  can be evaluated using Eq.~\eqref{eq:V} and the
value of $\mathcal{F}_\perp$ obtained. Since the recent $3\fb^{-1}$ of \lhcb
result~\cite{LHCb:2015dla}  does not provide branching fraction measurement for
the entire $q^2$ region we assume the theoretical values of
$\Gf$~\cite{Straub:2015ica,Horgan:2013pva} in addition to
$C_{10}$~\cite{Altmannshofer:2008dz}. The form factors $\mathcal{F}_\|$ and
$\mathcal{F}_0$ can then be determined from the fits to $\mathsf{P}_1$ and
$\mathsf{P}_2$  respectively, using Eq.~\eqref{eq:P_1-2}. Thus the conventional
form factors $A_1$ and $A_{12}$ can easily be estimated with the relation given
in Eqs.~\eqref{eq:A1} and \eqref{eq:A12}. In Table.~\ref{Table} we list the best
fit values with the $1\sigma$ uncertainties for the three form factors $V(q^2)$,
$A_1(q^2)$ and $A_{12}(q^2)$ for all the eight $q^2$ intervals. We also present
the standard deviation of the fit compared to the theoretical estimate from
Refs.~\cite{Straub:2015ica,Horgan:2013pva}.  While  sizable discrepancy is seen
for all the form factors especially in the regions $q^2<2.5~\gev^2$ and
$q^2>6~\gev^2$.  It is interesting to note the very significant discrepancy is
observed in the values of form factors $V$ and $A_1$ in bins $0.1\le q^2\le 0.98
~\gev^2$, $1.1\le q^2\le 2.5 ~\gev^2$, $6\le q^2\le 8 ~\gev^2$, $11.0\le q^2\le
12.5 ~\gev^2$ and $15\le q^2\le 17 ~\gev^2$. The lattice estimate of the
form factors currently does not include finite $\kstar$ width. This implies,
that the significance of the deviations can be lower if one includes the
unaccounted systematics due to the finite $\kstar$ width. We point out that
previous attempts to incorporate resonance contributions in theory has been done
by parametrically taking it’s effect in the Wilson coefficient $C_9$
\cite{Deshpande:1988bd,Kruger:1996cv}. However the accuracy of the 
form of resonance
parametrization does not alter our determination of form factors since, our
analysis is independent of $\widetilde{C}_9^{\sss\lambda}$ estimates.
$\widetilde{C}_9^{\sss\lambda}$ contributes only to $u_\lambda$'s and {\em the
ratios of form factors $\mathsf{P_1}$ and $\mathsf{P_2}$ do not get affected by
resonances.}  This is easily seen if we consider a situation where NP is absent
and all the parameters for resonances (strength, phase etc.) are known, the
observables calculated using Eqs.~\eqref{eq:Fperp}--\eqref{eq:A4} should agree
with the experimental measured observables. Thus the consistent set of
Eqs.~\eqref{eq:Fperp}--\eqref{eq:A4} must provide the same set of parameters
that we would have started with, as best fit solutions. In the absence of NP the
measured observables should result in the solutions for parameters matching with
SM values. Since $\mathsf{P_1}$ and $\mathsf{P_2}$  are unaffected by resonances
their best fit solutions also remain unaffected by it. Our best fit values of
$\mathsf{P_1}$ and $\mathsf{P_2}$ differ from the SM estimates and this
discrepancy cannot be accounted for by resonances.

To establish the above arguments we further undertake an
extensive study illustrated in Fig.~\ref{fig:P1-P2-Check}. We choose the region
$q^2> 11\,\gev^2$ as resonance effects can be dominant here and assume SM form
factor values of the observables from lattice calculations
\cite{Horgan:2013pva}. The solutions for $\mathsf{P_1}$ and $\mathsf{P_2}$ are
obtained using Eqs.~\eqref{eq:Fperp}--\eqref{eq:A4} for seven different $q^2$
points; $11\,\gev^2,~12\,\gev^2,~15\,\gev^2,~16\,\gev^2,~17\,\gev^2,~18\,\gev^2$
and $19\,\gev^2$.  The observables $\,F_\perp$, $\,F_L$, $\,A_4$, $\,\AFB^2$ and
$A_5^2$ are SM estimates calculated using lattice form factors. These seven
solutions of $\mathsf{P_1}$ and $\mathsf{P_2}$ are denoted by `star' symbols in
the corresponding plots. The black central line with gray band is the form
factor estimate (mean with $\pm 1\sigma$ error) of $\mathsf{P_1}$ and
$\mathsf{P_2}$. It can be seen that the set of
Eqs.~\eqref{eq:Fperp}--\eqref{eq:A4} are completely consistent with SM structure
and produces expected solutions. In case the solutions were completely
analytically obtained, the `stars'
should sit on the black curves. However the solutions for
hadronic parameters are very complicated and has been evaluated numerically,
resulting in small shifts that are visible. The blue
error bars are the solutions for $\mathsf{P_1}$ and $\mathsf{P_2}$ using the bin
average values of SM observables. It can be seen that as the
Eqs.~\eqref{eq:Fperp}--\eqref{eq:A4} are valid at each $q^2$ point, bin
averaging has induced some shifts in the solutions. However the results are in
agreement within $\pm1\sigma$ confidence level region.  To illustrate the effect
of resonances we have considered the parametrization from
Ref.~\cite{Kruger:1996cv}. The charmonium bound states $\jpsi(1S)$, $\psi(2S)$,
$\psi(3770)$, $\psi(4040)$, $\psi(4160)$ and $\psi(4415)$ are included in the
mentioned five observables. Interestingly, the change in the value of
observables including the resonances affected the solutions for
$\zeta$, $u_\perp$ and $u_\|$, however, solutions to $\mathsf{P_1}$ and $\mathsf{P_2}$ remained unaltered (upto second decimal place), hence, the solutions completely superimpose with the `stars' obtained without resonance contributions. We have also investigated the effect of resonances in the bin average where the observables are evaluated with lattice form factors including the above mentioned resonances and the solutions to $\mathsf{P_1}$ and $\mathsf{P_2}$ are shown in light blue bands for the three $q^2$ bins $11.0\le q^2\le 12.5 ~\gev^2$, $15\le q^2\le 17 ~\gev^2$ and $17\le q^2\le 19 ~\gev^2$. In this case the results with and without resonances do not completely superimpose however are quite consistent within $\pm 1\sigma$ error bars.
These solutions are also shown in Fig.~\ref{fig:P1P2} and \ref{fig:P1zeta}, in same light blue bands, for the relevant bins where resonance effect may in principle be significant.
The red error bars are the solutions for $\mathsf{P_1}$ and $\mathsf{P_2}$ obtained from data
(as discussed and highlighted in contours is Fig.~\ref{fig:P1P2}) that have been shown here again for convenience. We reiterate that effect
of resonances in  $\mathsf{P_1}$, $\mathsf{P_2}$ solutions is independent of the
parametrization choice as the solutions do not depend on Wilson coefficient
$\widetilde{C}_9^{\sss\lambda}$ and our conclusions derived for $\mathsf{P_1}$
and $\mathsf{P_2}$ parameters are unaffected by resonance effect. It is
justified that bin average can induce some errors in the solutions.
However, we have allowed a shift in the observable values (in Eq.~\ref{eq:chisq}
and Table.~\ref{sys-Table}) of more than the $1\sigma$ error for each observable
which hopefully is sufficient to compensate such effects.

It is important to note that in our analysis no hadronic estimates are used to
solve for the five parameters from exactly five measurements. Whereas, in other
approaches, when considering the same $B\to K^* \ell\ell$ mode all six form
factors, Wilson coefficients and non factorisable corrections based on
conservative estimations are needed. We compare $\mathsf{P}_1$ and
$\mathsf{P}_2$ obtained from experimental data alone, with the three form factor
$V$, $A_1$ and $A_{12}$ to which they are related as theoretical inputs. The
form-factors $T_1$, $T_2$ and $T_{23}$ are not used in this comparison. Thus,
our comparisons are different in nature and have reduced uncertainties, in terms
of number  of theoretical estimates. This may result in higher significance
level of deviation observed here.

The large $q^2$ region where the $\kstar$ has low-recoil energy has also been
studied~\cite{Grinstein:2004,Bobeth:2010wg} in a modified heavy quark effective
theory framework which is a model independent approach. In this limit the number
of independent hadronic form factors reduces to only three and one
finds~\cite{Das:2012kz} that $r_0=r_\|=r_\perp$ or equivalently
$u_0=u_\|=u_\perp$ must hold as long as non-factorizable charm loop
contributions are negligible. We find that this relation does not hold for
either of the bins $15\le q^2 \le 17~\gev^2$ or $17\le q^2 \le 19~\gev^2$. The
values of $u_0$, $u_\|$ and $u_\perp$ obtained from the fit with $\pm 1\sigma$
errors are listed in Table~\ref{Table-II}. We note that $u_\lambda$'s
receive  problematic resonance contribution coming from
$\widetilde{C}_9^{\sss\lambda}$. To address this issue we have introduced more
sytematics in measured observable than the one arising only from bin average
effect. We have checked our analysis by doubling the systematics of the
observables given in Table.~\ref{sys-Table} of Appendix.~\ref{sec:App2} for the
$q^2$ range $11\le q^2\le 12.5 ~\gev^2$ and $15\le q^2\le 17 ~\gev^2$ and our
results are stable with it. The actual significance of the deviations observed
here can be obtained with the detailed study of resonance systematics which is a
subject of an independent paper. However the significance level is evaluated by
conservatively adding systematics varying between 10\% -- 100\% in the
observables. The large discrepancies observed 
are equally hard to explain solely due to non-factorizable charm loop 
corrections and may be
additional evidence of physics beyond the SM.
\begin{table}[h]
\centering
\begin{tabular}{|c|c|c|c|}
\hline \hline $q^2$ range in $\gev^2$          & $u_0$ & $u_\|$ & $u_\perp$\\
\hline $15\le q^2\le 17$ & $0.000\pm 0.016$ & $0.013\pm0.153$  & $0.367\pm 0.025$   
\\ 
$17\le q^2\le 19$ & $0.166\pm 0.014$ & $0.000\pm 4.579$  & $0.260\pm 0.048$ \\ 
$15\le q^2\le 19$ & $0.120\pm 0.007$ & $0.004\pm 0.441$  & $0.244\pm 0.026$ \\
\hline 
    \hline 
  \end{tabular}
  \caption{ The values of $u_0$, $u_\|$ and $u_\perp$ obtained from fit to 
  $3\fb^{-1}$ of \lhcb data \cite{LHCb:2015dla}. In large $q^2$
  region~\cite{Grinstein:2004,Bobeth:2010wg} the equality $u_0=u_\|=u_\perp$  
  is expected
  to hold if non-factorizable charm loop contributions are negligible. The 
  errors in the
  value of $u_\|$ for the larger $q^2$ bin is unexpectedly large to draw any
  conclusions. Significant discrepancies which are too large to be solely
  due to non-factorizable charm loop corrections are observed between the values
  of $u_\perp$ and $u_0$ in both bins. } \label{Table-II}
\end{table}

\subsection{Testing relation between observables}
\label{subsec:obs} 

\begin{center}
\begin{figure*}[!bht]
 \begin{center}
	\includegraphics*[width=3in]{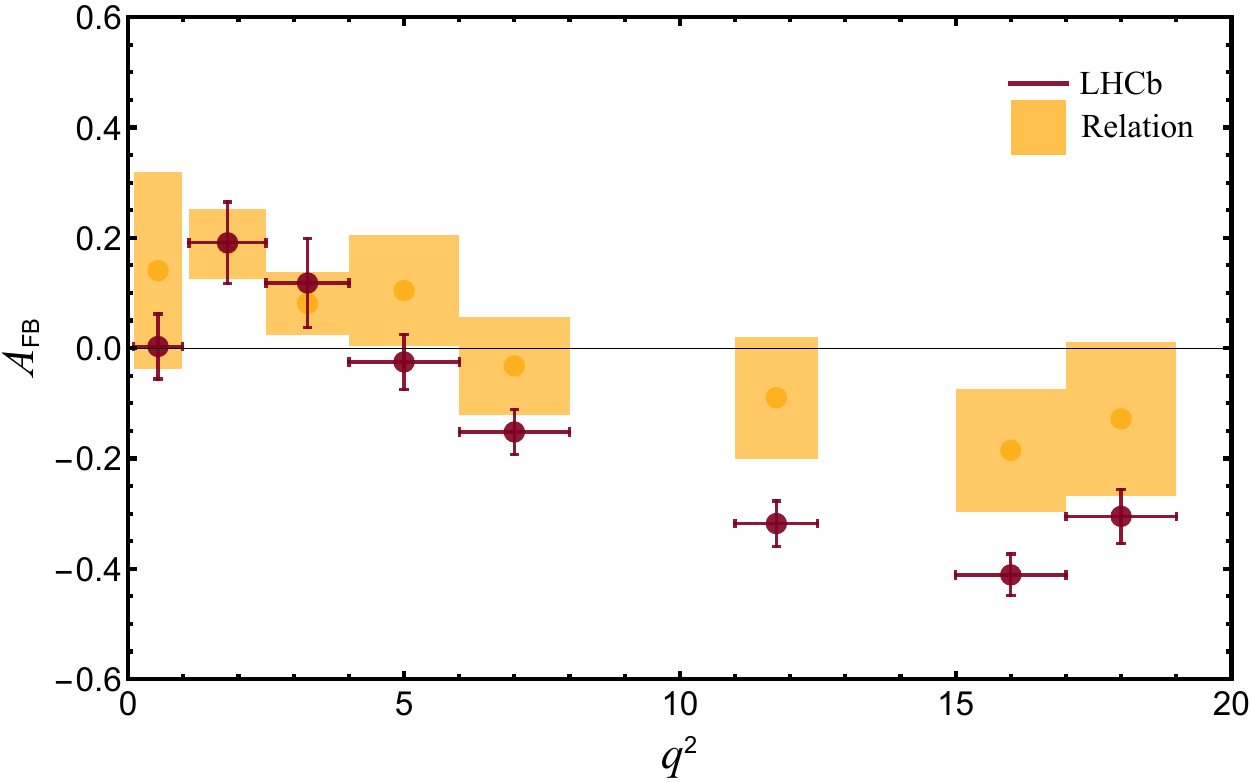}%
	\includegraphics*[width=3in]{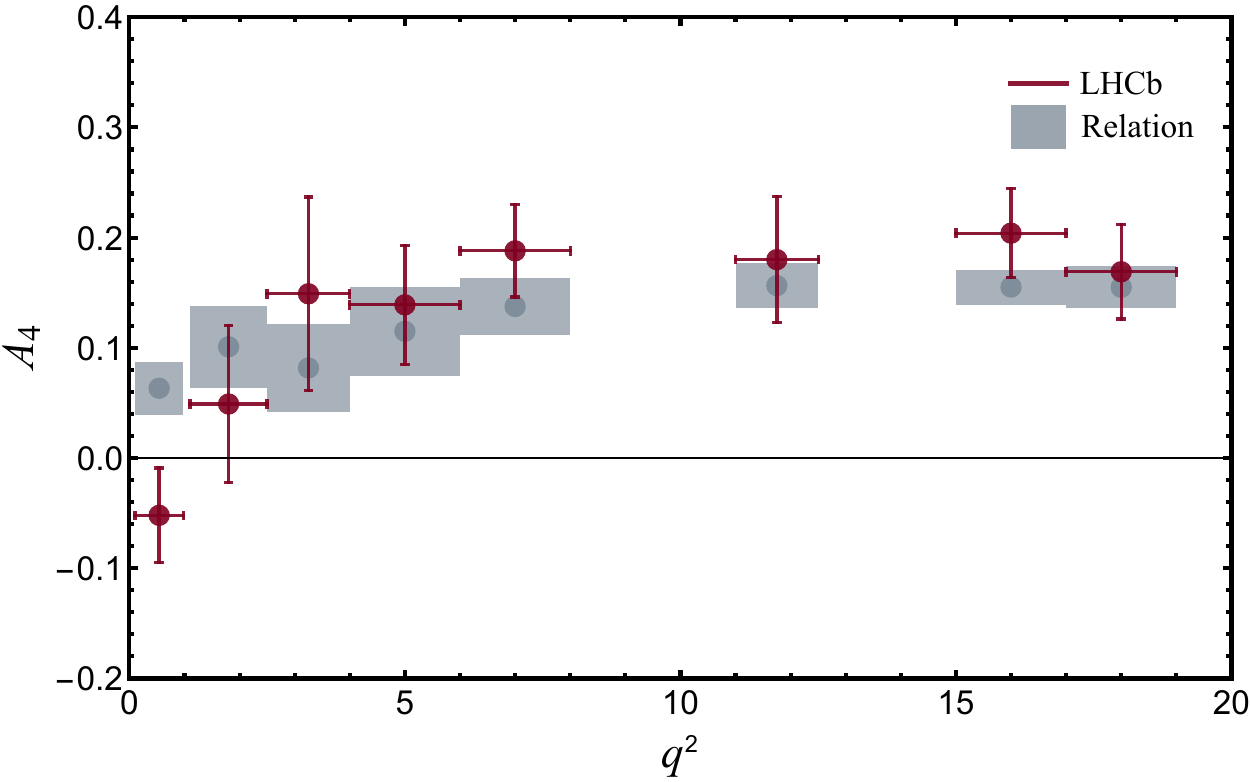}
	\includegraphics*[width=3in]{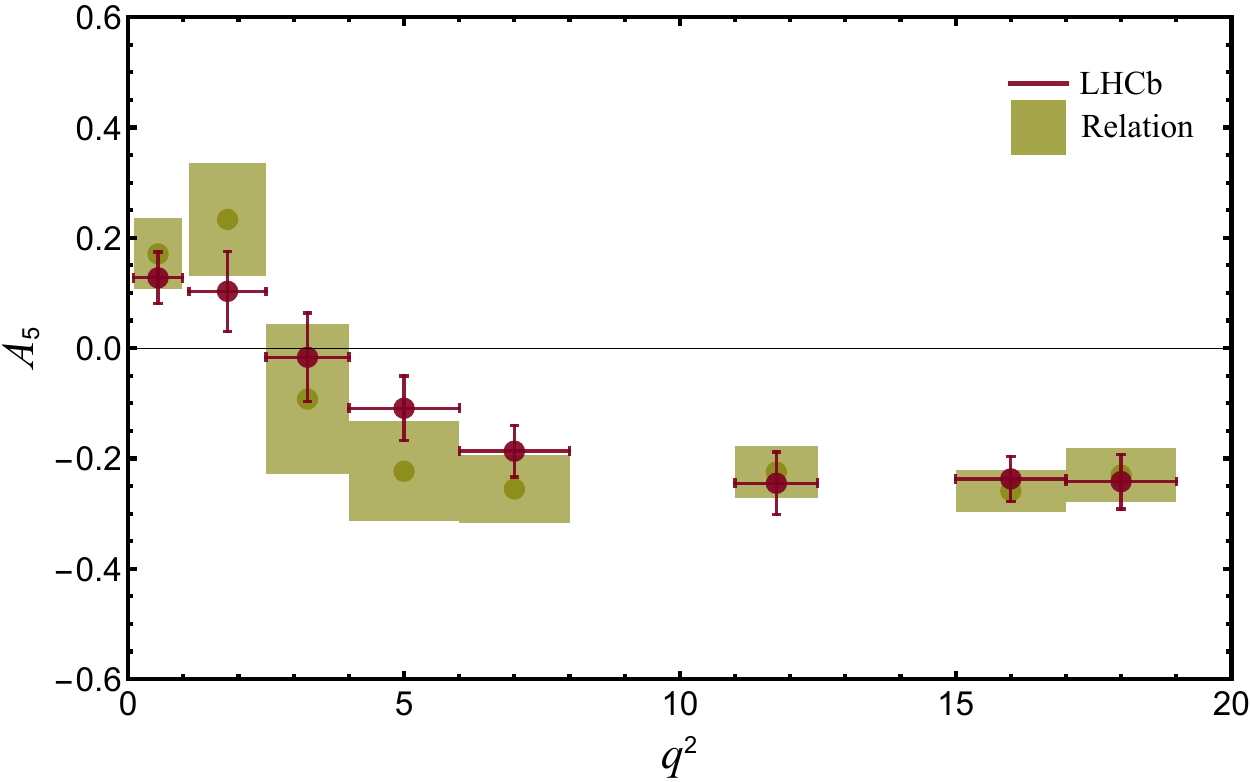}%
	\includegraphics*[width=3in]{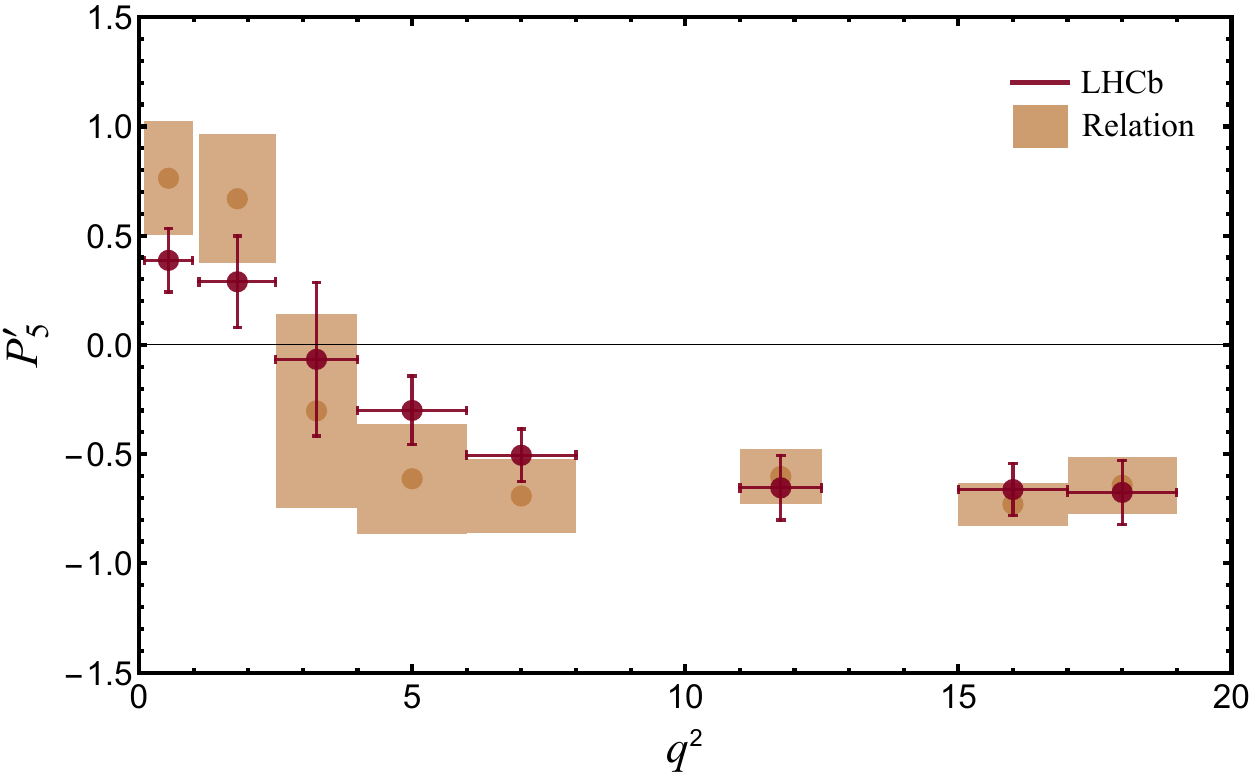}%
	\caption{(color online) The mean  values and $\pm 1\sigma$ uncertainty 
	bands for asymmetries $\AFB$, $A_4$, $A_5$ and $P_5^\prime$ 
	calculated using Eqs.~\eqref{eq:Obs-relationA4} -- 
	\eqref{eq:Obs-relationAFB}  are shown in yellow, gray, green and brown 
	bands, respectively. The error bars in red (dark) correspond to the \lhcb 
	measured \cite{LHCb:2015dla} central values and errors for each observable 
	for the respective $q^2$ bins. The predictions for the asymmetries are 
	obtained using the relations among observables which are independent of any 
	hadronic parameters and depend on experimental measurements of the other 
	observables remaining in the corresponding relations. Sizable discrepancies 
	are shown for $\AFB$ in $11.0\le q^2\le 12.5 ~\gev^2$ and $15\le q^2\le 17 
	~\gev^2$ bins and for $A_4$ in the range $0.1\le q^2\le 0.98~\gev^2$.
	We note that the relations (Eqs.~\eqref{eq:Obs-relationA4} -- 
	\eqref{eq:Obs-relationAFB}) remain valid except in the presence of NP 
	operators that result in modified angular distribution. Hence the presence of right-handed currents and any extra vector current such as $Z^\prime$ the relations will remain valid. } 
	\label{fig:obsAll}
 \end{center}
\end{figure*}
\end{center}

The relation between the observables for asymmetries $A_4$, $A_5$ and $\AFB$
given in Eqs.~\eqref{eq:Obs-relationA4} -- \eqref{eq:Obs-relationAFB}  can also
be tested using \lhcb data \cite{LHCb:2015dla}. In Fig.~\ref{fig:obsAll}, top
left panel, we compare theoretically calculated $\AFB$ mean values and $\pm
1\sigma$ errors (in yellow bands) with experimental measurements (red error
bars) for the respective $q^2$ bins. All observables in the r.h.s of
Eq.~\eqref{eq:Obs-relationAFB} (`relation' ) are assumed to be Gaussian
distributions in data and the predictions for $\AFB$ in yellow bands are
obtained using the expression of the `relation'. A very good agreement is
evident for most $q^2$ regions, however, for the ranges $11.0\le q^2\le 12.5
~\gev^2$ and $15\le q^2\le 17 ~\gev^2$ a deviation of $2.1\sigma$ and
$1.8\sigma$ is observed. Similarly `relation' for $A_4$ in
Eq.~\eqref{eq:Obs-relationA4} results in a very good agreement except for
showing a discrepancy of $2.3\sigma$ only in the $0.1\le q^2\le 0.98~\gev^2$
bin, in right top panel of Fig.~\ref{fig:obsAll}. The disagreement in the value
of $\AFB$ and $A_4$ in some $q^2$ bins indicates that there is no set of form
factors and Wilson coefficients which can explain $\AFB$ and $A_4$ completely.
Observables $A_5$ or equivalently $P_5^\prime$~\cite{DescotesGenon:2012zf} are
found to be in complete agreement i.e. within about $\pm 1\sigma$ deviation for
all $q^2$ bins as shown in the two lower panels of Fig.~\ref{fig:obsAll}. The
solutions for $A_5$ and $\AFB$ have ambiguities. We chose the ambiguity for
which the chi-squared deviations are the least. Our conclusions have no bearing
on and do not rule out the observation made by \lhcb in observable $P_5^\prime$  in
Refs.~\cite{LHCb:2015dla,Aaij:2013qta}. The predictions of observable
$P_5^\prime$ derived from the relation is a signal of consistency of \lhcb
results. We note that the relation remains valid except in the presence of NP
operators that result in modified new angular distribution. Hence we do not
expect to see the discrepancy observed by \lhcb 
\cite{Aaij:2013qta} if right-handed currents or
extra vector current such as $Z'$ contributes to the decay. The discrepancy observed by \lhcb depends on the comparison with
model based calculation of form factors. Whereas, the predictions of these
asymmetries made in this paper, are independent of any form factor values and
depend purely on the gauge structure of SM. If the model dependent calculations
of form factors are correct, signal of new physics may well be indicated in the
bins suggested by Ref.~\cite{Aaij:2013qta}. We find that \lhcb data indicates
yet another independent discrepancy.

\section{Conclusion} 
\label{sec:conclusion} 

In conclusion, we have used the $3\invfb$ of \lhcb data to determine some hadronic
parameters governing the decay $B\to \kstar \ell^+\ell^-$ assuming contributions
from SM alone. We obtain the values of the form factors $V(q^2)$,
$A_1(q^2)$ and $A_{12}(q^2)$ that are used to describe the matrix element
$\braket{\kstar}{\bar{s}\gamma^{\mu}P_L b}{B}$ directly from data. Very 
significant deviations are seen for the form factors $V$ and $A_1$ especially in the regions
$q^2<2.5~\gev^2$ and $q^2>6~\gev^2$.
We point out that the presence of resonances in data can induce more systematic uncertainties in the fits. However in the view of absence of such a existence of resonances in $B\to \kstar \ell^+ \ell^-$ data, we emphasize that the significant deviations observed in the form factor values can not be completely explained by resonances and non-factorizable contributions. 
We would like to point out that there exist  major differences between the global fit
approaches \cite{Descotes-Genon:2015uva} to study the anomalies in $b \to s$ transitions and the approach adopted in our work. Our work relies only on $B\to K^* \ell\ell$ decay mode,
whereas, global fit techniques incorporate various decay modes and hence either 
use LCSR, Lattice based estimates of form factors or treat form-factors as 
parameters in the fit procedure. The number of inputs and fitted parameters 
differ making a number by number comparison of the 
different approaches difficult.
Furthermore due to the absence of accurate estimates of non factorisable corrections, the global fit techniques rely on some conservative estimations of these corrections. However, the formalism we have developed parametrizes such corrections and the conclusions drawn here are independent of non factorisable estimates. 
These are perhaps the reasons why we find larger significance. However,
qualitatively we don't see a significant disagreement with the other
approaches as we do observe $\sim 3\sigma$ discrepancy in
$\mathsf{P_1}-\mathsf{P_2}$ plane in $q^2$ region $[6-8]\gev^2$ where observable $P_5^\prime$ also deviates by $2.7\sigma$ from its SM prediction.

Further, a relation between form factors 
expected
to hold in the large $q^2$ region as long as non-factorizable charm loop
contributions are negligible, seems to fail. 
Finally, the relation between observables also indicates some deviations in the 
same regions where the form factors were found to disagree. The 
forward-backward asymmetry $\AFB$ deviates in the $q^2>11~\gev^2$ region, where 
as $A_4$  differs in the region $q^2\le 0.98~\gev^2$. 
As the systematic error arises from the experimental measurements of observables in
terms of  binned dilepton invariant mass are accounted, the magnitude of discrepancies
observed would be hard to accommodate either as systematics from long distance resonance contributions or possible corrections to theoretical estimates. 
All these features can be understood if there are other unaccounted for
operators contributing to the decay mode. 
In view of this, we speculate that these
deviations are likely to be a signature of physics beyond SM.

\acknowledgments
We are indebted to Tom Browder and thank him for several suggestions and
discussions. We also thank  J. Martin Camalich, Hai-Yang Cheng, N. G Deshpande, 
Jim Libby and Arjun Menon for discussions. R.S. thanks Hai-Yang Cheng and 
Institute of Physics, Academia Sinica, Taipei, Taiwan for hospitality during 
final stages of manuscript preparation.

\appendix

\section{Complex contribution \boldmath$\varepsilon_\lambda$ estimates from data}
\label{sec:App1}

In Ref.~\cite{Mandal:2014kma} it was shown that the complex contributions 
$\varepsilon_\lambda$ to the
amplitude of the decay mode $B\to\kstar \ell^+ \ell^-$, 
can be taken into consideration. 
$\varepsilon_\lambda$ can be solved in terms of iterative solutions proportional
to the observables $A_7$, $A_8$, $A_9$ and a form factor ratio $\mathsf{P}_1$.
The expressions for all the three $\varepsilon_\lambda$'s are shown in 
Eqs.~(76)--(78) 
of Ref.~\cite{Mandal:2014kma}. They are reproduced here for convenience.
\begin{align}
\label{eq:eps_perp}
\varepsilon_\perp&=\frac{\sqrt{2}\pi\Gf}{(r_0\!-\!r_\|)\mathcal{F}_{\!\perp}} 
\Bigg[\frac{A_9 \mathsf{P_1}}{3\sqrt{2}}+\frac{A_8 \mathsf{P_2}}{4}
-\frac{A_7 \mathsf{P_1}\mathsf{P_2}r_\perp}{3\pi C_{10} }\Bigg], 
\\[2ex]
\label{eq:eps_parallel}
\varepsilon_\|&=\frac{\sqrt{2}\pi\Gf}{(r_0\!-\!r_\|)\mathcal{F}_{\!\perp}} 
\Bigg[\frac{A_9 r_0}{3\sqrt{2} r_\perp}+\frac{A_8 \mathsf{P_2}r_\|}
{4 \mathsf{P_1}r_\perp}-\frac{A_7\mathsf{P_2} r_\|}{3\pi C_{10}}
 \Bigg], \\[2ex]
\label{eq:eps_0}
\varepsilon_0&=\frac{\sqrt{2}\pi\Gf}{(r_0\!-\!r_\|)\mathcal{F}_{\!\perp}} 
\Bigg[\frac{A_9\mathsf{P_1}r_0}{3\sqrt{2} \mathsf{P_2}r_\perp}+\frac{A_8 
r_\|} {4 r_\perp} -\frac{A_7 \mathsf{P_1} r_0}{3\pi C_{10} }\Bigg]. 
\end{align}
A point to be noted as explained in detail in Ref.~\cite{Mandal:2014kma}, is 
that the $(\varepsilon_\lambda/\Gf^{\nicefrac{1}{2}})$'s  are completely 
expressed in terms of
observables and the form factor ratio $\mathsf{P_1}$. However, these solutions
are essentially iterative, since the $r_\lambda$'s and $C_{10}$ are
derived in terms of the primed observables that depend on $\varepsilon_\lambda$.
If $(\varepsilon_\lambda/\Gf^{\nicefrac{1}{2}})$ are small as should be 
expected, accurate solutions for them can be found with a few iterations.
In Ref.~\cite{Mandal:2014kma} the variation of $\varepsilon_\lambda$ with 
$\mathsf{P}_1$ was 
studied for $1\invfb$ LHC data and it was found that the solutions are not 
sensitive to the value of $\mathsf{P}_1$.

We generate a set of events for every bin, with each event consisting of
randomly chosen values drawn from Gaussian distributions 
generated for each of the observables $F_L$, $F_\perp$, $A_4$,
$A_5$, $\AFB$, $A_7$, $A_8$ and $A_9$. The distributions are generated using 
experimental results from Ref.~\cite{LHCb:2015dla}, with
the  experimentally measured value as mean and the uncertainty as standard
deviation.
\begin{center}
\begin{figure}[!h]
 \begin{center}
	\includegraphics*[width=03.4in]{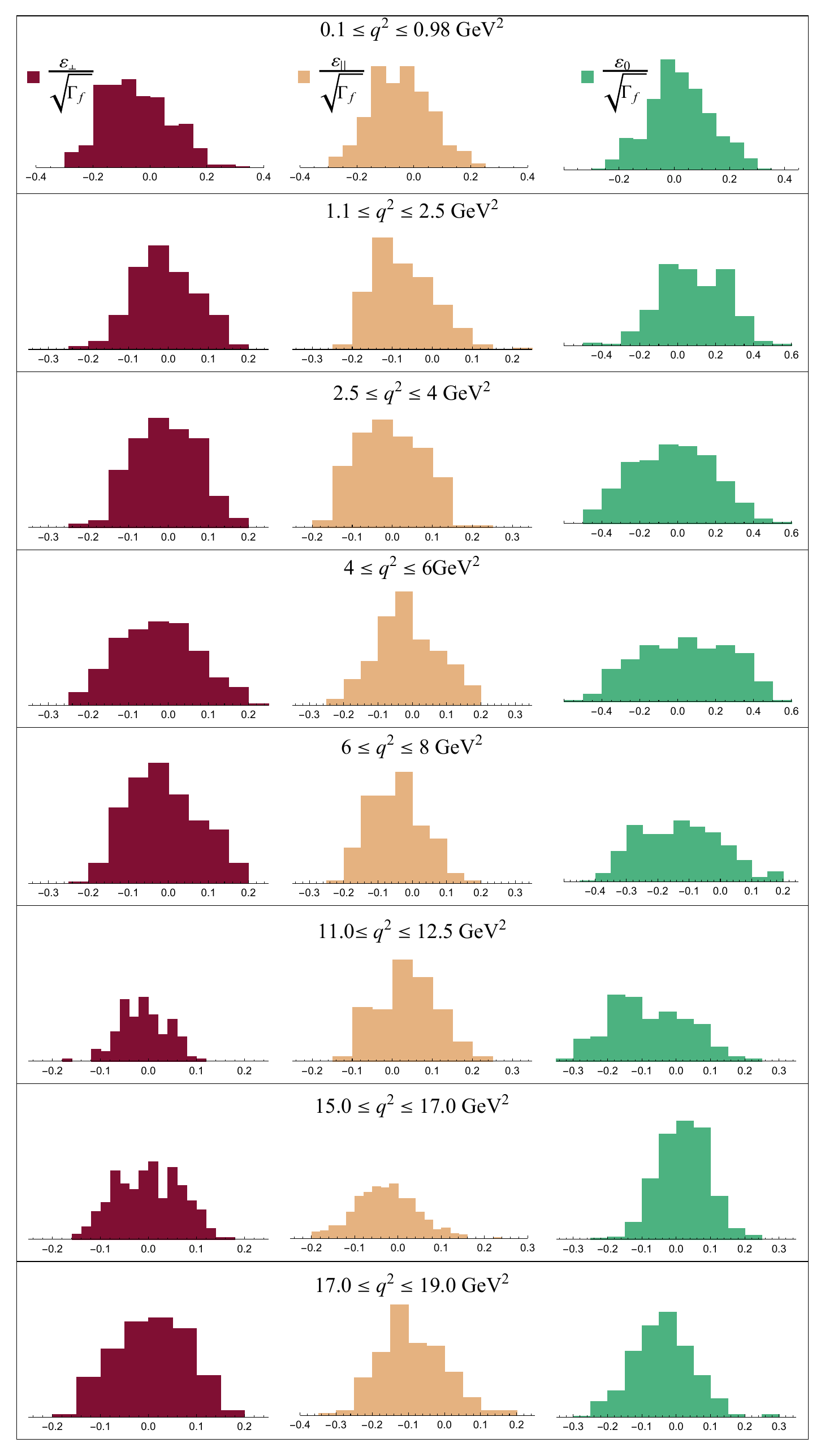}%
	 \caption{(color online) The solutions for $\varepsilon_\perp/\sqrt{\Gf}$, 
	 $\varepsilon_\|/\sqrt{\Gf}$ and $\varepsilon_0/\sqrt{\Gf}$ using 
	 distributions for first through eighth $q^2$ bins are depicted
	 in red (dark), light brown (lightest) and green respectively.
	 All the $\varepsilon_\lambda/\sqrt{\Gf}$'s are consistent with zero.} 
	 \label{fig:epsilon}
\end{center}
\end{figure}
\end{center}

\begin{table}[h]
\centering
  \begin{tabular}{ c | c| c| c| }
    \hline \hline
    $q^2$ in $\gev^2$ & $\varepsilon_\perp/\sqrt{\Gf}$ & $\varepsilon_\|/\sqrt{\Gf}$ & $\varepsilon_0/\sqrt{\Gf}$  \\ \hline
    $0.1\!\! \le \!\! q^2 \!\! \le \!\! 0.98$ & $-0.048 \pm0.116$ & $-0.047 \pm 0.103$&$ 0.020 \pm \! 0.111$ \\ 
    $1.1\!\! \le \!\! q^2 \!\! \le \!\! 2.5$  & $-0.010\pm 0.078$ & $-0.010\pm 0.078$ & $0.078\pm \! 0.172$ \\ 
    $2.5\!\! \le \!\! q^2 \!\! \le \!\! 4.0 $ & $-0.009\pm 0.079$ & $-0.008\pm 0.080$ & $-0.025\pm \! 0.212$\\ 
    $4.0\!\! \le \!\! q^2 \!\! \le \!\! 6.0 $ & $-0.026 \pm 0.097$& $0.014\pm 0.093$ & $0.032 \pm \! 0.234$\\ 
    $6.0\!\! \le \!\! q^2 \!\! \le \!\! 8.0 $ & $-0.011 \pm 0.088$& $-0.046\pm 0.078$& $-0.132 \pm \! 0.129$ \\ 
    $11.0\!\! \le \!\! q^2 \!\! \le \!\! 12.5$& $-0.011 \pm 0.050$& $0.038\pm 0.074$& $-0.078 \pm \! 0.114$ \\ 
    $15.0\!\! \le \!\! q^2 \!\! \le \!\! 17.0$& $-0.000\pm 0.067$ & $-0.027\pm 0.071 $& $0.020\pm \! 0.072$\\ 
    $17.0\!\! \le \!\! q^2 \!\! \le \!\! 19.0$& $0.006\pm 0.076$ & $-0.090\pm 0.090$ & $ -0.040\pm \! 0.088$\\ \hline
    \hline 
  \end{tabular}
  \caption{The $\varepsilon_\lambda/\sqrt{\Gf}$ mean values with $\pm 1\sigma$ errors from Fig.~\ref{fig:epsilon}}
\label{eps-Table}
\end{table}

$\varepsilon_\lambda$ are solved iteratively for every set of observables.  We
find converged iterative solutions for $\varepsilon_\lambda/\sqrt{\Gf}$ for each
set of observables with the histograms shown in Fig.~\ref{fig:epsilon}. The red
(dark), light brown (lightest) and green histograms denote the solutions for
$\varepsilon_\perp/\sqrt{\Gf}$, $\varepsilon_\|/\sqrt{\Gf}$ and
$\varepsilon_0/\sqrt{\Gf}$ respectively for all the eight bins with $q^2$ range
$(0.1-0.98)~\text{GeV}^2$, $(1.1-2.5)~\gev^2$, $(2.5-4.0)~ \gev^2$,
$(4-6)~\gev^2$, $(6-8)~\gev^2$, $(11.0-12.5)~\gev^2$, $(15-17)~\gev^2$ and
$(17-19)~\gev^2$.

We have also quoted the mean and $\pm 1\sigma$ errors for each
$\varepsilon_\lambda/\sqrt{\Gf}$ in Table.~\ref{eps-Table} calculated from the
distributions shown in Fig.~\ref{fig:epsilon}. It can be easily seen that all the mean
values of $\varepsilon_\lambda/\sqrt{\Gf}$ are consistent with zero. From
Eqs.~(37)--(40) of Ref.~\cite{Mandal:2014kma}, the contributions from imaginary
part of the amplitude to the observables $F_L$, $F_\|$, $F_\perp$ and $A_4$ are
quadratic in the corresponding $\varepsilon_\lambda/\sqrt{\Gf}$ and thus are
negligible.

\section{Systematic uncertainty evaluation}
\label{sec:App2}

\begin{center}
\begin{figure*}[hbtp]
 \begin{center}
	\includegraphics*[width=2.25in]{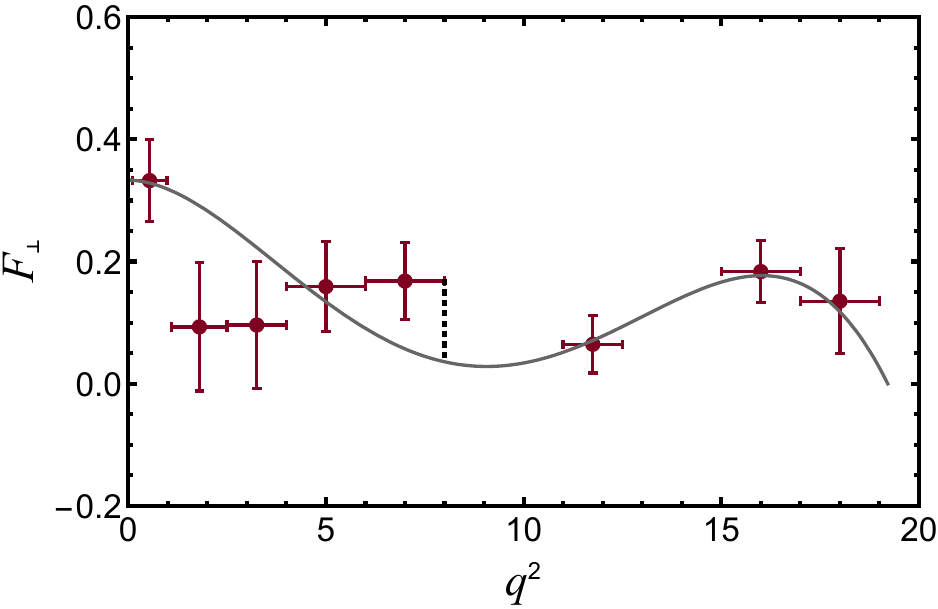}%
	\includegraphics*[width=2.2in]{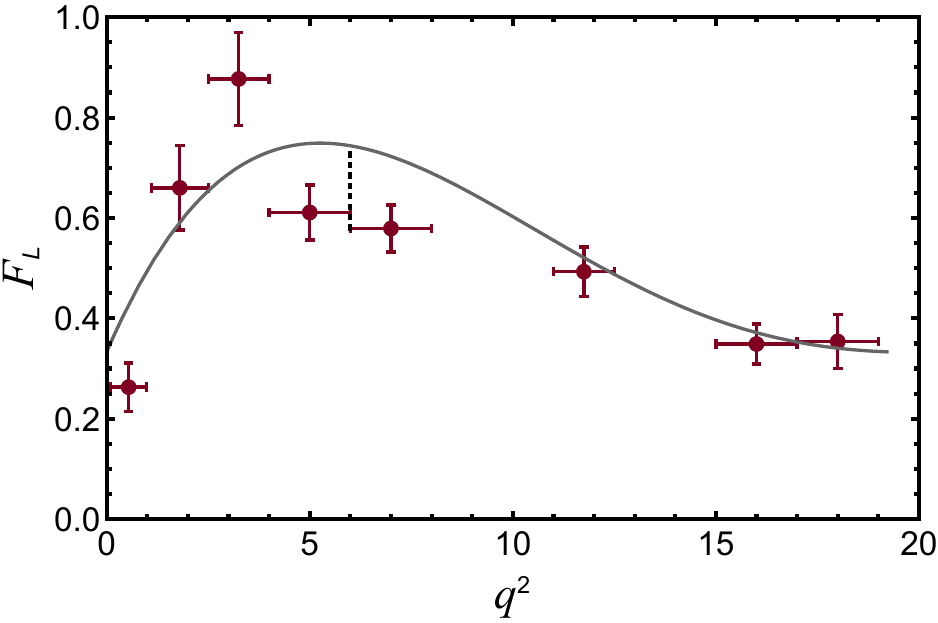}
	\includegraphics*[width=2.25in]{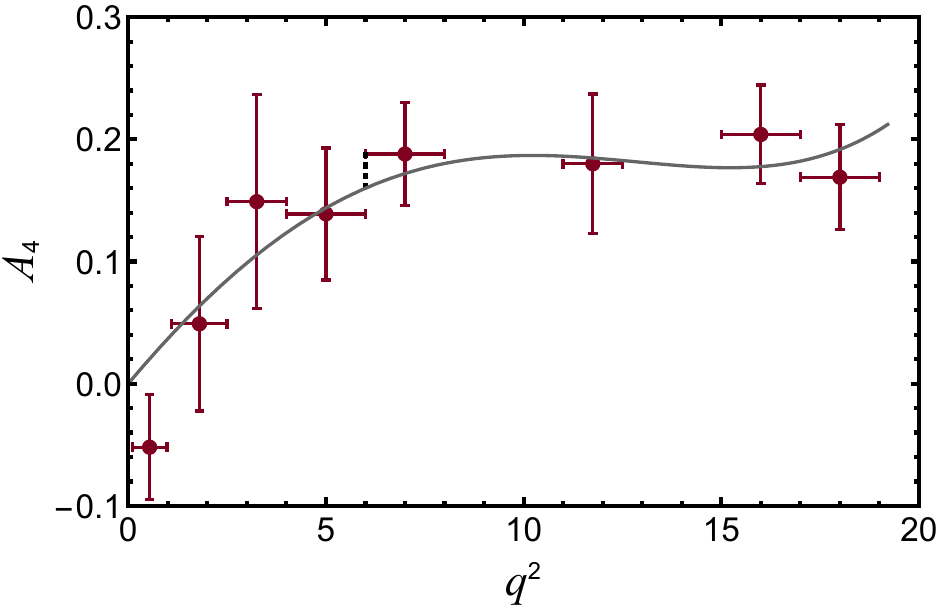}
	\includegraphics*[width=2.25in]{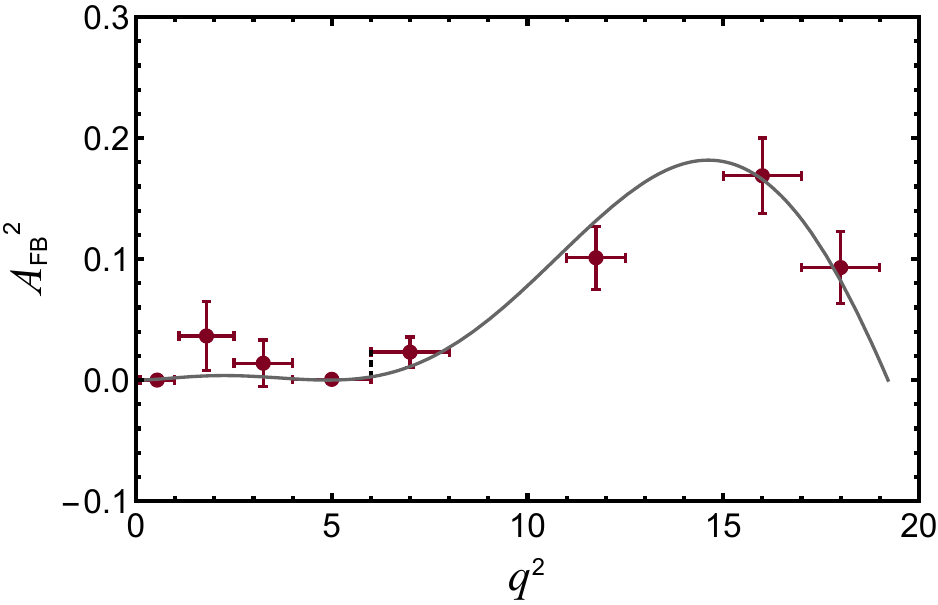}%
	\includegraphics*[width=2.25in]{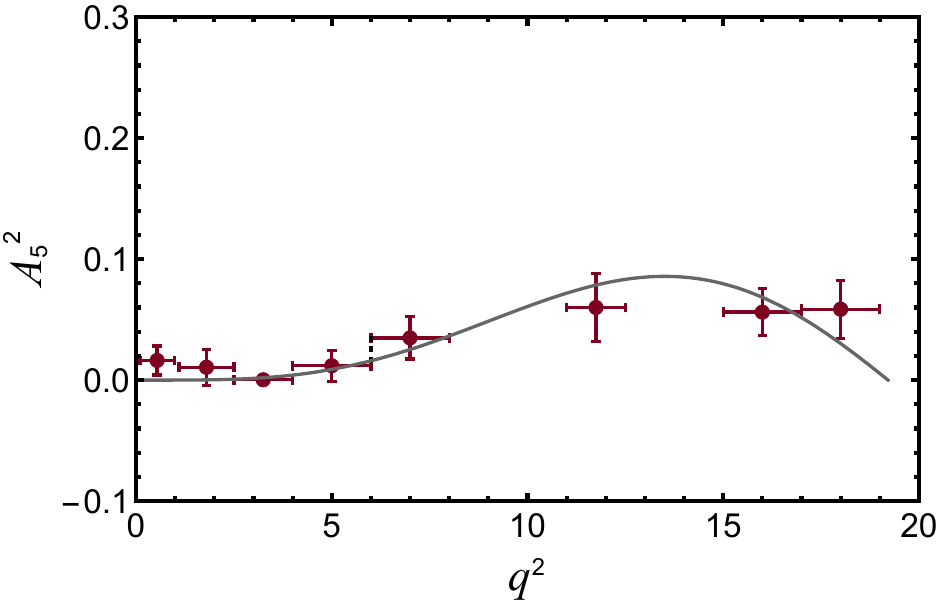}
	\caption{(color online). The procedure to calculate systematic errors are shown for 
	observables $\,F_\perp$, $\,F_L$, $\,A_4$, $\,\AFB^2$ and $A_5^2$, respectively. 
	The red error bars are \lhcb measurements and gray curves represent best fitted 
	polynomial in $q^2$ for 14 bin \lhcb data \cite{LHCb:2015dla}. The black dashed lines denote the 
	maximum deviation of bin average central value of the observables with the $q^2$ function for bin $6 \le q^2 \le 8~ 
	\gev^2$. The length of these black lines are denoted by $F_\perp^s$, $F_L^s$,
	$A_4^s$, 
	${\AFB^2}^{\!\!\!\!\!s}$  and ${A_5^2}^{\!\,s}$, respectively. Similar lines can be 
	drawn for other $q^2$ bins also and the values of systematic errors are given in 
	Table.~\ref{sys-Table} for all observables.
} \label{fig:sys}
\end{center}
\end{figure*}
\end{center}

\begin{table}[h!]
\centering
  \begin{tabular}{ c | c| c| c| c| c }
    \hline \hline
    $q^2$ range in $\gev^2$ & $F_\perp^s$ & $F_L^s$  & $A_4^s$ & ${\AFB^2}^{\!\!\!\!\!s}$ & ${A_5^2}^{\!\,s}$  \\ 
    \hline
    $0.1\le q^2 \le 0.98$ & $0.014 $  & $0.230$ & $0.088 $ & $0.002$ & $0.016$\\ 
    $1.1\le q^2 \le 2.5$  & $0.223 $  & $0.151 $& $0.036$ & $0.034$  & $0.010$\\ 
    $2.5\le q^2 \le 4.0 $ & $0.164$  & $0.223$& $0.064$   & $0.013$  & $0.004$\\  
    $4.0\le q^2 \le 6.0 $ & $0.069$  & $0.138$& $0.021$   & $0.002$  & $0.008$\\ 
    $6.0\le q^2 \le 8.0 $ & $0.132$  & $0.165$& $0.028$   & $0.020$  & $0.019$\\ 
    $11.0\le q^2 \le 12.5$& $0.029$  & $0.063$& $0.006$   & $0.051$  & $0.023$\\ 
    $15.0\le q^2 \le 17.0$& $0.019$  & $0.048$& $0.027$   & $0.036$  & $0.023$\\ 
    $17.0\le q^2 \le 19.0$& $0.109$  & $0.020$& $0.039$   & $0.077$  & $0.053$\\ 
    \hline \hline 
  \end{tabular}
\caption{The systematic uncertainties for each observables $\,F_\perp$, $\,F_L$, $\,A_4$, $\,\AFB^2$ and $A_5^2$ are shown. The values denote magnitude of maximum deviation of the bin average central value with the fitted $q^2$ polynomial within every $q^2$ bin. }
\label{sys-Table}
\end{table}

We discuss the evaluation of systematic uncertainties arising mainly due to bin average effect of observables. As written in Eq.~\eqref{eq:shift}, the shift $\mathcal{O}^{s}$ in each observable is calculated for each $q^2$ bin, by considering the maximum deviation of the bin average value of the observable $\mathcal{O}$ from a fitted $q^2$ polynomial of entire range. It is highlighted in Fig.~\ref{fig:sys} where red error bars are \lhcb measurements and gray curves represent best fitted polynomial in $q^2$ for 14 bin \lhcb data. We use 14 bin measurement (based on the method of moments \cite{Beaujean:2015xea}) from \lhcb to fit the polynomial in $q^2$, rather than the 8 bin data set as it provides more information to determine the shape of the polynomial for entire $q^2$ region. The black dashed line denotes the maximum deviation of bin average central value of the observable with the $q^2$ function for the region $6 \le q^2 \le 8~ \gev^2$ and $\mathcal{O}^{s}$ is the length of the line for observable $\mathcal{O}$. Similar technique is applied for other $q^2$ bins also and the values of systematic errors are given in Table.~\ref{sys-Table} for all observables.

It should be noted that as discussed in Sec.~\ref{sec:numerics} finite lepton mass 
can affect the analysis in the first two $q^2$ region namely $q^2 \le 2.5 
~\gev^2$ and in the absence of a measurement of asymmetries $A_{10}$ and $A_{11}$ \cite{Mandal:2014kma} we have 
to rely on some hadronic estimates. This in principle may cause more 
uncertainties and we took a conservative approach by considering  two times the 
$\mathcal{O}^{s}$ values for all observables given in Table.~\ref{sys-Table} for 
the two bins $0.1\le q^2 \le 0.98~\gev^2$ and $1.1\le q^2 \le 2.5~\gev^2$. 

We emphasize that resonances in our analysis will only affect the fitted 
function in $q^2$, which in turn will induce more systematic uncertainties to 
the observables. We have checked the $\chi^2$ fit (in Sub-sec.~\ref{subsec:ff}) by increasing the 
systematic uncertainties two times of the values given in Table.~\ref{sys-Table} for the regions $11 \le q^2 \le 12.5~ \gev^2$ and $15 
\le q^2 \le 17~ \gev^2$ and our results are stable with it. However a detailed study of resonance systematics on this decay mode is currently going on and will be a subject of an independent paper itself.

\end{document}